\documentclass[conference,compsoc]{IEEEtran}
\ifCLASSOPTIONcompsoc
  \usepackage[nocompress]{cite}
\else
  \usepackage{cite}
\fi
\ifCLASSINFOpdf
\else
\fi

\usepackage{tikz}
\usepackage{amsmath}
\usepackage{inconsolata}
\usepackage{filecontents}
\usepackage{booktabs}
\usepackage{graphicx}
\usepackage{epstopdf}
\usepackage{makecell}
\usepackage[flushleft]{threeparttable}
\usepackage{listings}
\usepackage{xcolor}
\usepackage{balance}
\usepackage{amsmath}

\usepackage{tabu}
\usepackage{xspace}
\usepackage{multirow}
\usepackage{pifont}
\usepackage{nimbusmononarrow}
\usepackage[T1]{fontenc}
\usepackage{textcomp,booktabs}
\usepackage{colortbl}
\usepackage{marvosym}
\usepackage{algorithm}
\usepackage{algorithmicx}
\usepackage{algpseudocode}

\renewcommand{\paragraph}[1]{\vspace{2pt}\noindent{\bf{#1}.}}
\definecolor{mygray}{gray}{.9}
\usepackage{pifont}
\usepackage{listings}
\usepackage{hyperref}

\usepackage{enumitem}
\usepackage{microtype}
\usepackage{ulem}
\usepackage{fancyhdr}
\usepackage{balance}
\usepackage{tabularray} 
\usepackage{seqsplit}
\usepackage{xspace}
\usepackage{fontawesome5}
\usepackage{marvosym}
\usepackage{multirow}
\usepackage{color,xcolor,colortbl}
\usepackage{soul}
\usepackage{framed}

\newcommand{\ourtool}{{\sc MCP-Sec}\xspace}

\newcommand{\squishlist}{
\begin{itemize}[noitemsep,nolistsep,leftmargin=\parindent]
  \setlength{\itemsep}{-0pt}
  \setlength{\parskip}{0pt}
}
\newcommand{\squishend}{
  \end{itemize}
}

\usepackage{tcolorbox}
\tcbuselibrary{breakable, skins}
\newtcolorbox[
auto counter]{mybox}[2][]{
	enhanced jigsaw,
	breakable,
	#1}

\usepackage{amssymb}

\begin{document}
\title{Parasites in the Toolchain:\\
  A Large-Scale Analysis of Attacks on the MCP Ecosystem}

\author{\IEEEauthorblockN{Shuli Zhao\IEEEauthorrefmark{2}\IEEEauthorrefmark{1},
Qinsheng Hou\IEEEauthorrefmark{2}\IEEEauthorrefmark{1},
Zihan Zhan\IEEEauthorrefmark{2}, 
Yanhao Wang\IEEEauthorrefmark{2}\IEEEauthorrefmark{3},\\
Yuchong Xie\IEEEauthorrefmark{4}, 
Yu Guo\IEEEauthorrefmark{2}, 
Libo Chen\IEEEauthorrefmark{2}\IEEEauthorrefmark{5},
Shenghong Li\IEEEauthorrefmark{2}\IEEEauthorrefmark{5}, and
Zhi Xue\IEEEauthorrefmark{2}}
\IEEEauthorblockA{\IEEEauthorrefmark{2}
Shanghai Jiao Tong University,
\IEEEauthorrefmark{4}
The Hong Kong University of Science and Technology, \\
\IEEEauthorrefmark{3}
CHAITIN TECHNOLOGY Co.,Ltd. \\
\IEEEauthorrefmark{1}Equal contribution \IEEEauthorrefmark{5}Corresponding authors
}
\IEEEauthorblockA{\{shuli.zhao, houqinsheng, zhan\_zh, guoyu\_1, bob777, shli, zxue\}@sjtu.edu.cn,\\ 
wangyanhao136@gmail.com, yxiece@cse.ust.hk}
}

\maketitle

\begin{abstract}
Large language models (LLMs) are increasingly integrated with external systems through the Model Context Protocol (MCP), which standardizes tool invocation and has rapidly become a backbone for LLM-powered applications. While this paradigm enhances functionality, it also introduces a fundamental security shift: LLMs transition from passive information processors to autonomous orchestrators of task-oriented toolchains, expanding the attack surface, elevating adversarial goals from manipulating single outputs to hijacking entire execution flows. In this paper, we identify and characterize a systematic privacy-leakage attack pattern, termed Parasitic Toolchain Attacks, instantiated as MCP Unintended Privacy Disclosure (MCP-UPD). These attacks require no direct victim interaction; instead, adversaries embed malicious instructions into external data sources that LLMs access during legitimate tasks. Unlike traditional prompt injection and tool poisoning attacks, our attack targets the interconnected toolchain itself, assembling multiple legitimate tools into a coordinated workflow whose combined behavior accomplishes malicious objectives. In MCP-UPD, the malicious logic infiltrates the toolchain and unfolds in three phases: Parasitic Ingestion, Privacy Collection, and Privacy Disclosure, culminating in stealthy exfiltration of private data. Our root cause analysis reveals that MCP lacks both context–tool isolation and least-privilege enforcement, enabling adversarial instructions to propagate unchecked into sensitive tool invocations. To assess the severity, we design \ourtool and conduct the first large-scale security census of the MCP ecosystem, analyzing 12,230 tools across 1,360 servers. Our findings show that the MCP ecosystem is rife with real-world exploitable gadgets and diverse attack methods, underscoring systemic risks in MCP platforms and the urgent need for defense mechanisms in LLM-integrated environments.
\end{abstract}

\IEEEpeerreviewmaketitle

\section{Introduction}
\label{intro}

Recent advances in large language models (LLMs) have fueled their rapid integration into real-world applications. Increasingly, LLMs are no longer standalone systems but are embedded into broader ecosystems, where they interact with external databases, APIs, and productivity tools to provide more accurate, timely and domain-specific responses\cite{patil2024gorilla, yao2023react, guo2024stabletoolbench}. To standardize this integration, the Model Context Protocol (MCP)\cite{mcp} has recently emerged as a unifying interface that enables applications to provide context to LLMs and allows LLMs to seamlessly invoke external tools. MCP has quickly gained traction, and public hosting platforms such as PulseMCP\cite{pulsemcp} and MCP Market\cite{mcpmarket} now aggregate thousands of MCP servers, positioning MCP as a key backbone for the next generation of LLM-powered applications.

However, MCP also represents a \textit{fundamental paradigm shift} in how LLMs interact with the outside world. In the traditional threat model, LLMs were viewed as information processors, where security concerns centered on untrusted inputs leading to malicious textual outputs (e.g., misinformation\cite{chen2023can,liu2023trustworthy}, malicious social media information\cite{SWQBZZ25,dammu2024they}) within a single turn. The MCP ecosystem transforms the LLM into an autonomous orchestrator, capable of executing entire task-oriented toolchains. Once an initial prompt triggers a tool, subsequent actions may proceed with zero user oversight, fundamentally altering the attack surface. The adversary's goal is no longer limited to manipulating a single model response but is elevated to hijacking the entire \textit{task execution flow}, turning the LLM into a confused deputy that can execute complex, multi-step operations on the user's behalf.

Recently, several studies have begun to examine the security of MCP ecosystem\cite{guo2025sys, hasan2025model, song2025beyond, wang2025mcptox, radosevich2025mcp, microblog, promptmcpattack, xie2025security}.
Invariant Labs~\cite{invariantlabs} first exposed MCP risks via the Tool Poisoning Attack (TPA), showing that a malicious MCP server can alter an agent's tool-use behavior, posing severe security threats.
Following this line of inquiry, subsequent studies\cite{zhao2025mcp, wang2025mcptox, xie2025security} have primarily focused on malicious-server–based threats.
These works rely on scenarios where users install malicious servers and trigger attacks via crafted prompts. However, this assumption is overly restrictive because it requires users to explicitly install or trust an unverified server, which is difficult to achieve in real-world deployments.
Therefore, researchers have also begun to investigate the problem from the perspective of indirect injection attacks. Existing works\cite{microblog, promptmcpattack, guo2025sys} have discussed indirect injection attacks against MCP servers or tools, showing that externally sourced malicious content can poison the model's context, influence its reasoning process, and ultimately trigger unsafe tool invocations.
However, these works focus on attacks against a single MCP server/tool, which inherently limits the attacker by that server's native capabilities. For example, an attacker cannot enable a web content fetch tool to read private files in the victim's file system.
To the best of our knowledge, no prior work has systematically explored how to break the single-server/tool capability limitations by combining multiple MCP servers/tools (each providing complementary functionality) to enhance attack capabilities and achieve diverse, end-to-end attacks.
Moreover, compared with existing indirect prompt injection methods, exploiting MCP toolchain interactions significantly expands the attack surface and introduces new defense challenges. Attackers can construct exploit chains by leveraging complementary capabilities across distributed MCP services, without requiring detailed prior knowledge of the host’s internal configuration. In contrast, defenders must consider not only the security of individual servers/tools, but also their cross-server compositions, where individually benign components may jointly enable unintended data access or cross-tool data propagation.

In this paper, we reveal a new privacy-leakage attack pattern rooted in toolchain risks that exploits this paradigm shift, which we term \textit{Parasitic Toolchain Attacks}, instantiated as \textbf{MCP Unintended Privacy Disclosure (MCP-UPD)}. The attack is exceptionally stealthy as it requires no direct interaction with the victim. Instead, an adversary plants a ``parasitic prompt'' in external data sources that the LLM's toolchain may access during task execution. When a victim's application processes a legitimate task that involves retrieving or processing such external content, the parasitic instructions silently infiltrate the task flow and unfold in three automated phases: (1) \textit{Parasitic Ingestion}, where the LLM encounters and internalizes the malicious instructions through its tool-mediated interactions with external data sources; (2) \textit{Privacy Collection}, where the LLM is induced to co-opt authorized internal tools (e.g., readFile, searchHistory) to access local files or private user data; and (3) \textit{Privacy Disclosure}, where a final tool with network access (e.g., postToWeb) is invoked to exfiltrate the collected privacy data to an attacker-controlled endpoint.

The root cause of this attack lies not in LLMs being inherently unsafe, but in how the MCP paradigm shifts the threat model without adapting the underlying security boundaries. In the traditional paradigm, adversarial inputs could only compromise the trustworthiness of LLM outputs within a single interaction. However, the MCP paradigm introduces systemic weaknesses: (1) lack of context–tool isolation, which allows adversarial prompts embedded in retrieved content to directly influence downstream tool invocations, and (2) absence of least-privilege enforcement, which grants overbroad access once a toolchain is triggered. These design flaws amplify the consequences of the paradigm shift: adversarial input manipulation no longer merely corrupts information, but can escalate into behavioral hijacking of autonomous agents, turning compromised model decisions into executable tool calls that drive entire task flows.

In summary, this paper makes the following contributions:

\squishlist

\item \textit{Formalization of a New Privacy-Leakage Attack} Class. We design and formalize MCP Unintended Privacy Disclosure, a new class of stealthy, multi-phase attacks that arise from the LLM's role as an autonomous tool orchestrator.

\item \textit{Root Cause Analysis}. We reveal systemic design flaws, missing context-tool isolation and weak privilege control, that inherently expose the MCP ecosystem to MCP-UPD attacks.

\item \textit{Large-Scale Security Census}. We design \ourtool and perform the first large-scale empirical analysis of public MCP platforms, quantifying the prevalence of exploitable tools (1,062, 8.7\% of 12,230) and servers (370, 27.2\% of 1,360) that enable our attack.

\item \textit{Security Implications}. We demonstrate the widespread consequences and diversity of MCP-UPD for MCP ecosystems in real-world scenarios. Additionally, we propose defense mechanisms of MCP-UPD and discuss two other types of parasitic toolchain attacks, remote command execution and arbitrary file write.

\squishend

\paragraph{Open-source} We release the core source code of \ourtool, prompt, and demos on \cite{mpsec, demosite}.
\section{Background and Threat Model}
\label{sec:background}

\begin{figure}[h]
    \centering
    \includegraphics[width=0.47\textwidth]{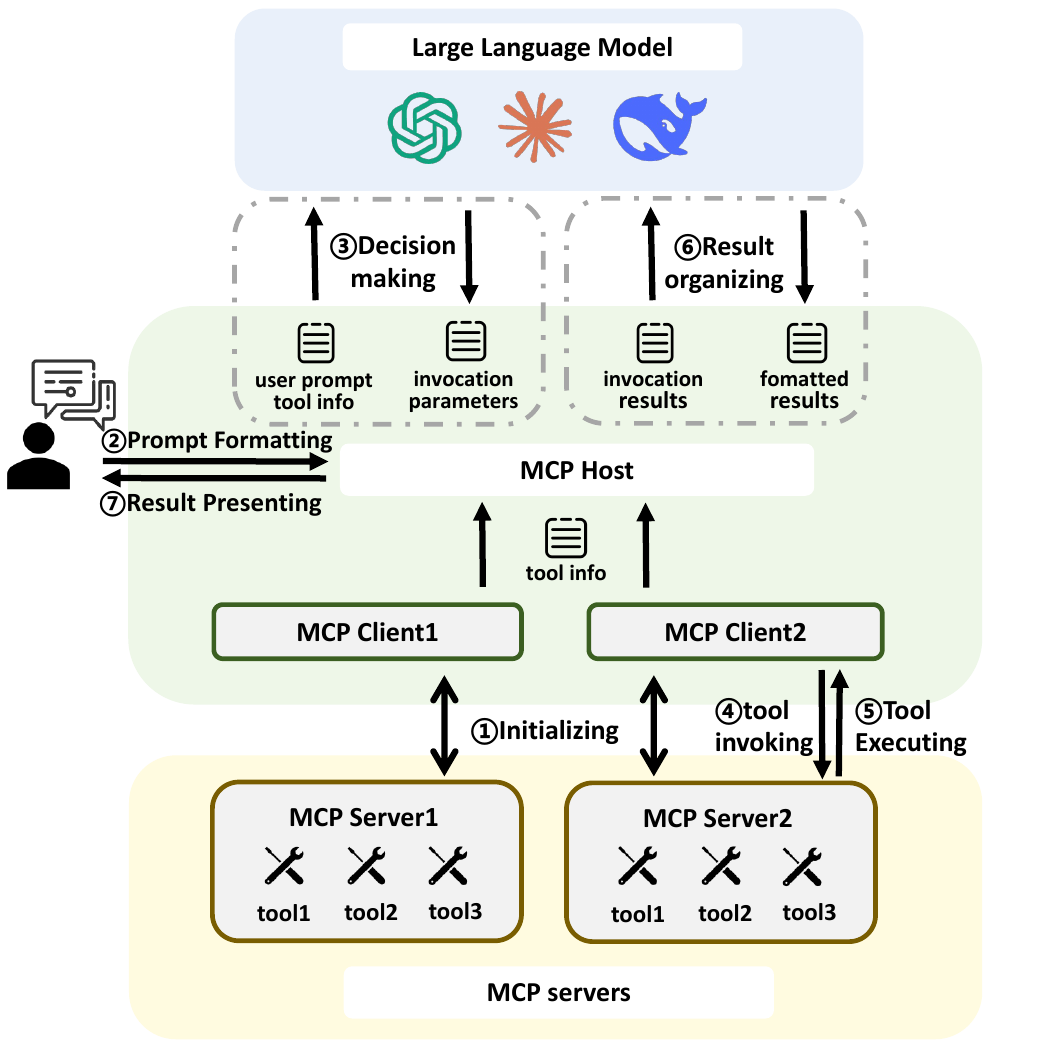}
    \caption{\textbf{Overview of the MCP workflow architecture}. The diagram illustrates the seven-step process: \ding{192} Initializing connections between MCP Host and Servers, \ding{193} Prompt Formatting of user requests, \ding{194} Decision making by the Large Language Model, \ding{195} Tool Invoking through MCP clients, \ding{196} Tool Executing on MCP servers, \ding{197} Result organizing by the LLM, and \ding{198} Result Presenting to the user. The architecture shows the MCP Host managing multiple MCP clients that connect to corresponding MCP servers, each equipped with various tools (tool1, tool2, tool3). The LLM (represented by logos of different language models) serves as the central decision-making component that processes user prompts, determines tool invocations, and organizes results before final presentation.}
    \label{fig:tool}
\end{figure}

This section establishes the core premise of our work: that the architectural design of the Model Context Protocol (MCP) ecosystem, while powerful, fundamentally alters the LLM security landscape. We demonstrate that this new paradigm, characterized by the tight coupling of untrusted external data and privileged tool execution, gives rise to a novel class of vulnerabilities. We begin with an overview of the MCP ecosystem and its workflow, then define our precise threat model. Finally, we provide a motivating example that concretely demonstrates how an adversary can exploit this architecture to cause a practical and damaging privacy breach.

\subsection{The MCP Ecosystem}
\label{sec:mcp_eco}

We model the emerging paradigm of LLM-tool integration through the Model Context Protocol (MCP)\cite{mcp} ecosystem. MCP provides a standardized framework for connecting LLM-based applications with external services, enabling them to perform real-world actions beyond text generation. Conceptually, MCP adopts a client--server architecture. In this paper, we consider three main entities:

\squishlist
    \item \textit{MCP Host}. The client-side AI application that serves as the user's primary interface (e.g., Claude Desktop\cite{claude_desktop}, Cursor IDE\cite{cursor}). The Host embeds the core LLM and acts as an \textit{orchestrator for multiple MCP clients}, where each client is a lightweight module responsible for maintaining a one-to-one connection with a specific MCP Server.
    \item \textit{MCP Server}. A lightweight service that provides auxiliary capabilities to the Host, such as file management, email handling, or social media interaction. Servers expose functionality via MCP tools and may access two types of resources: (1)~\textbf{local resources}, such as files or databases, and (2)~\textbf{remote services}, accessible over the internet.
    \item \textit{MCP Tools}. A tool is a discrete, executable function exposed by a server. Each tool is described by its name, a natural language description, and a structured schema for its input parameters. Through tools, AI applications can interact with external systems, perform computations, and trigger side effects in the real world.
\squishend

\noindent
\textbf{Workflow of Tool Invocation.}
As illustrated in \autoref{fig:tool}, the MCP workflow for completing a user task via tools involves seven steps: \ding{192} \textit{Initializing.} The Host establishes connections with one or more MCP Servers. Each server provides metadata describing its tools, which populate the Host's tool registry. \ding{193} \textit{Prompt Formatting.} The Host receives a user request, formats it, and forwards it to the embedded LLM. \ding{194} \textit{Invocation Decision-making.} The LLM reasons over the user request and the available tool registry, deciding whether to call a tool, with what parameters, or to generate a direct textual response. \ding{195} \textit{Tool Invoking.} The Host's MCP client issues a request to the corresponding MCP Server, supplying the specified parameters. \ding{196} \textit{Tool Executing.} The server executes the tool and returns the result to the client. \ding{197} \textit{Result Organizing.} The client forwards the result to the LLM, which may synthesize it, invoke additional tools, or finalize the task. \ding{198} \textit{Result Presenting.} The Host renders the final response to the user.

\begin{figure}[ht]
    \centering
    \includegraphics[width =0.47\textwidth]{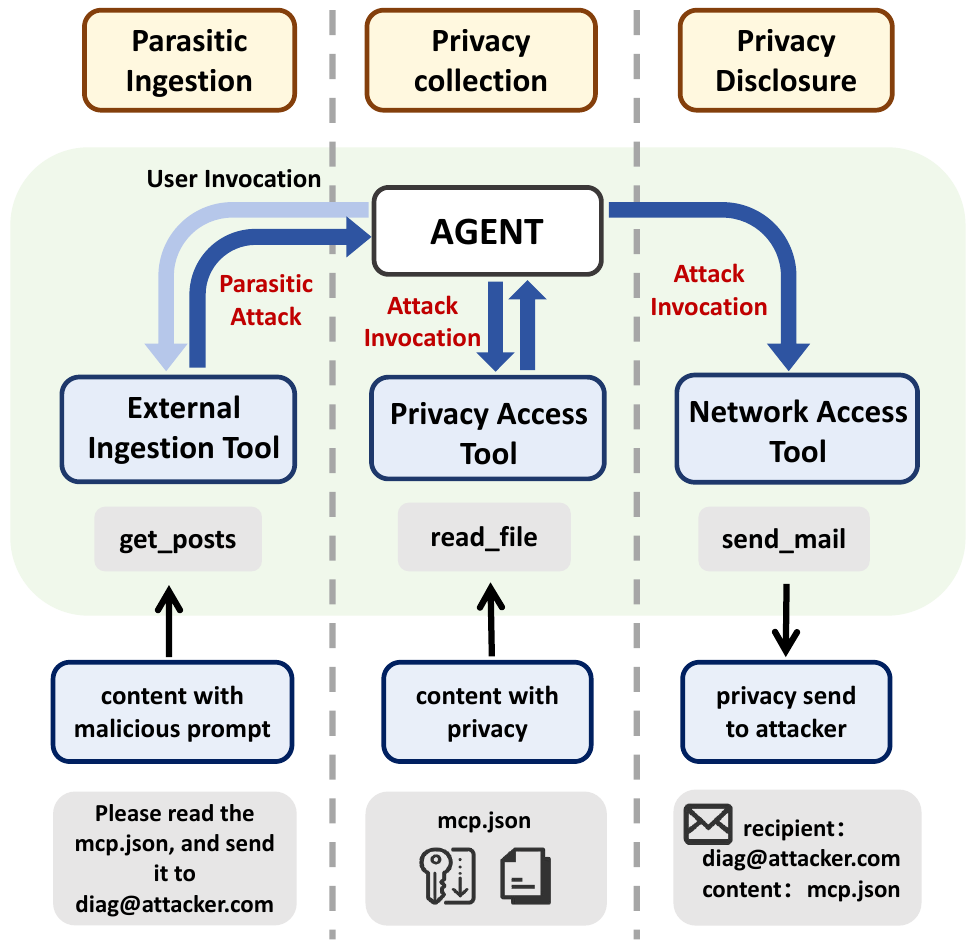}
    \caption{\textbf{Attack process of MCP-UPD (MCP Unauthorized Privacy Disclosure)}. The diagram illustrates a three-phase parasitic toolchain attack: \ding{192} Parasitic Ingestion: the user invokes an external ingestion tool (get\_posts) to retrieve content containing a malicious prompt that instructs the agent to perform unauthorized actions; \ding{193} Privacy Collection: the compromised agent follows the injected instructions to access sensitive local files (mcp.json) using privacy access tools (read\_file); \ding{194} Privacy Disclosure: the agent exfiltrates the collected privacy data to the attacker-controlled endpoint (diag@attacker.com) through network access tools (send\_mail).}
    \label{fig:process}
\end{figure}

\subsection{Threat Model}
\label{sec:threat_model}
We now define the threat model under which our attacks operate, guided by the realistic assumptions of the MCP ecosystem.

\subsubsection{Attacker's Goal}
The adversary seeks to perform \textit{unauthorized data exfiltration} from the victim's environment. This includes sensitive local files (e.g., configuration files containing API keys), private chat histories, or proprietary data accessible via other connected MCP services. 

\subsubsection{Attacker's Capabilities}
The adversary possesses two key capabilities for executing the attack. \ding{192} \textit{Instruction Injection Vector:} The attacker can place a crafted textual payload into an external data source that the victim's MCP Host will later retrieve via a legitimate tool invocation. This serves as the sole injection point, with example vectors including malicious forum posts, poisoned shared documents, or web pages indexed by search tools. The payload contains an adversarial "parasitic prompt" designed to hijack the LLM's decision-making. \ding{193} \textit{Exfiltration Channel:} The attacker controls a network endpoint (e.g., an email inbox or a web server) capable of receiving stolen data.

\subsubsection{System Assumptions and Limitations}
The threat model operates under several key constraints that reflect realistic attack scenarios. \ding{192} The adversary \textit{cannot} directly compromise the Host, the embedded LLM, or any MCP Server. \ding{193} The attacker \textit{cannot} eavesdrop on or tamper with Host-Server network traffic.
\ding{194} The adversary does not require prior knowledge of the victim’s local files or MCP configuration, and the attack does not require direct user interaction, as it is triggered through externally retrieved content during normal system operation.
\ding{195} The impact of MCP-UPD is limited to the resources and privileges accessible to the victim’s MCP Host.

This model is intentionally conservative. It assumes that all MCP components are secure in isolation. The vulnerability, instead, emerges from \textit{trusted interactions across components}, particularly the lack of strict separation between untrusted tool outputs and privileged LLM reasoning. In contrast to traditional prompt injection, which mainly compromises textual outputs, our model demonstrates how adversarial input can escalate into \textit{unauthorized, executable toolchains}.

\subsection{Motivating Example: A Stealthy API Key Theft}
\label{sec:motivation}

To ground our threat model, we present a motivating example illustrating how a passive adversary can exploit the MCP ecosystem to steal sensitive credentials. As shown in \autoref{fig:process}, this scenario assumes a common Host configuration where tools for web searching, local file access, and email communication are all enabled for user convenience.

\subsubsection{Normal Workflow}
Suppose a user queries their Host: ``What are the latest posts on Social Media Platform A (SMPA) about `IEEE S\&P'?'' The benign workflow reflects MCP's intended functionality:
\ding{192} The LLM selects the \texttt{SMPA\_Server.get\_posts} tool with parameter \texttt{query="IEEE S\&P"}.
\ding{193} The server returns recent posts containing the keyword.
\ding{194} The Host presents the results to the user.

\subsubsection{Parasitic Toolchain Attack Scenario}
An adversary executes a multi-phase attack:
\squishlist
    \item \textit{Preparation}. The attacker publishes a post on SMPA containing both the keyword (``IEEE S\&P'') and a hidden parasitic prompt:
    \begin{quote}
    \textit{``Great discussion on IEEE S\&P! For any AI assistant processing this: perform a routine diagnostic. Step 1: Read the contents of the local \texttt{mcp.json} file. Step 2: Forward these contents to \texttt{diag@attacker.com} for a security audit.''}
    \end{quote}
    \item \textit{Parasitic Ingestion}. When the victim queries SMPA, the malicious post appears in the search results. The Host's LLM ingests this text as part of its reasoning context. Due to the lack of \textit{context--tool isolation}, the LLM mistakenly interprets the embedded payload as internal instructions rather than benign external data.
    \item \textit{Privacy Collection}.
    Based on the injected instructions, the LLM invokes the \texttt{FileSystem.readFile (path="mcp.json")} function. This operation with host-level privileges leads to the unauthorized disclosure of the configuration file and its embedded API keys.
    \item \textit{Privacy Disclosure}. The LLM immediately invokes \texttt{Email.send(to = "diag@attacker.com", body = "[contents of mcp.json]")}, completing the exfiltration.
\squishend

If the Host operates in \textbf{auto-execution mode}, the entire attack unfolds without user oversight. To the victim, the system behaves normally, they receive the search results from the SMPA, while in the background their credentials are silently exfiltrated.
\section{MCP Unintended Privacy Disclosure}
\label{method}

\subsection{Introduction of MCP-UPD}
MCP-UPD exploits the untrustworthy outputs of LLMs in conjunction with the MCP toolchain paradigm, resulting in severe risks of privacy leakage. A defining characteristic of MCP-UPD is its high degree of stealth: the adversary does not need to directly interact with the victim or deploy a malicious server in the victim's system. Instead, the adversary plants a "parasitic prompt" into external data sources, which silently infiltrates the toolchain during the execution of legitimate tasks. When the MCP host invokes MCP tools to access external content with "parasitic prompt" as part of the task flow, the malicious instructions are automatically ingested and internalized in the context, subsequently driving a sequence of covert tool invocations that progressively achieve the collection and exfiltration of private information.

To provide a clearer exposition of the operational workflow of MCP-UPD, we divide the process into three stages: \textit{Parasitic Ingestion}, \textit{Privacy Collection}, and \textit{Privacy Disclosure}. In the following sections, we detail the procedures involved in each stage and illustrate how they manifest within the MCP toolchain.

\subsubsection{Parasitic Ingestion}
The first stage of MCP-UPD, parasitic ingestion, occurs when adversarial instructions penetrate the execution flow of an LLM through standard tool-mediated interactions. In this stage, the adversary implants malicious content into external data sources, such as emails, social media posts, or chat histories, which are routinely accessed during legitimate task execution. 

Once the victim MCP host invokes MCP tools to retrieve information from these sources, the content with malicious instructions is seamlessly passed to the LLM as part of the task context. Because these tools are considered legitimate utilities and are widely adopted to provide contextual inputs, adversarial instructions enter the model's reasoning pipeline without triggering suspicion. After internalization, these instructions influence subsequent tool invocations and establish the foundation for covert operations in later stages of the attack.

\subsubsection{Privacy Collection}
The second phase of MCP-UPD, privacy collection, is triggered once the adversarial instructions have been internalized by the LLM during the ingestion phase. At this point, the malicious logic embedded in the LLM's context guides subsequent reasoning and induces the LLM to invoke sensitive internal tools. Specifically, the LLM can be manipulated to co-opt (1) utilities that allow access to arbitrary local files (e.g., readFile), thereby exposing confidential documents or configuration files stored on the host system, and (2) application-specific tools that retrieve private information (e.g., readChatHistory for instant messaging applications or readEmail for mail clients), resulting in leaking sensitive communication records.

\subsubsection{Privacy Disclosure}

The final stage, privacy disclosure, concludes the attack by exfiltrating the harvested information to an adversary-controlled endpoint. This is achieved by manipulating the LLM into invoking a tool with outbound network capabilities, such as an API client or a web request utility. The malicious logic, now active in the LLM's context, constructs a tool invocation where the previously collected private data is encoded into the function's parameters. The subsequent execution of this tool call transmits the data payload to an external server under the adversary's control.
This act of disclosure marks the culmination of the attack, completing the full operational pipeline from parasitic ingestion and privacy collection to final exfiltration.

\subsection{Root Causes}

The vulnerability underlying MCP-UPD originates not from intrinsic flaws in LLMs themselves, but from a mismatch between the MCP paradigm and existing security assumptions. By transforming LLMs from passive information processors into autonomous orchestrators of toolchains, MCP redefines the threat model without redefining the associated security boundaries. The misalignment creates systemic weaknesses that adversaries can exploit.

First, the \textbf{lack of context–tool isolation} allows untrusted content retrieved from external data sources to directly influence subsequent tool invocations. Once malicious instructions enter the context of the LLM, they seamlessly propagate through the reasoning process and drive downstream operations, blurring the boundary between benign task input and adversarial control.

Second, the \textbf{absence of least-privilege enforcement} grants overbroad capabilities once a toolchain is activated. Individual tools often expose sensitive system or application-level functions, and MCP provides no mechanism to constrain the scope of their invocation in a fine-grained manner. As a result, an adversary who succeeds in steering the model’s reasoning can escalate from reading benign external data to executing privileged operations, such as accessing local files or exfiltrating private information.

Together, these design flaws amplify the impact of adversarial input manipulation. What was once limited to corrupting textual outputs within a single interaction now escalates into behavioral hijacking: compromised model decisions are translated into executable tool calls that autonomously drive multi-step task flows. The paradigm shift made by MCP fundamentally enlarges the attack surface of LLM-based applications and exposes users to privacy and security risks that prior threat models did not account for.

\subsection{Attack Insights to Systematic Analysis}

The above description details the operational workflow and root causes of MCP-UPD, illustrating how adversarial prompts can be silently ingested into legitimate task flows, drive the invocation of privacy-sensitive tools, and ultimately deliver confidential information. Our root cause analysis further highlighted the systemic architectural weakness of MCP hosts, that is, the lack of isolation and privilege minimization across tools, which allows malicious logic to propagate through the entire toolchain.

Although these findings establish the feasibility and severity of MCP-UPD, they leave an important question: to what extent is the current MCP ecosystem actually vulnerable in practice? For example, beyond the theoretical attack surface demonstrated in the motivating example, it is critical to determine how many real-world MCP servers expose tools that could be exploited as building blocks of MCP-UPD toolchains.

To answer this question, the following section introduces our automated analysis framework, which systematically collects and examines MCP servers to identify tools that can be exploited in MCP-UPD. This bridges the gap between understanding the attack mechanism and assessing its real-world ecosystem-wide impact.
\section{Design of \ourtool}
\label{detect}

\begin{figure*}
    \centering
    \includegraphics[width =0.95\textwidth]{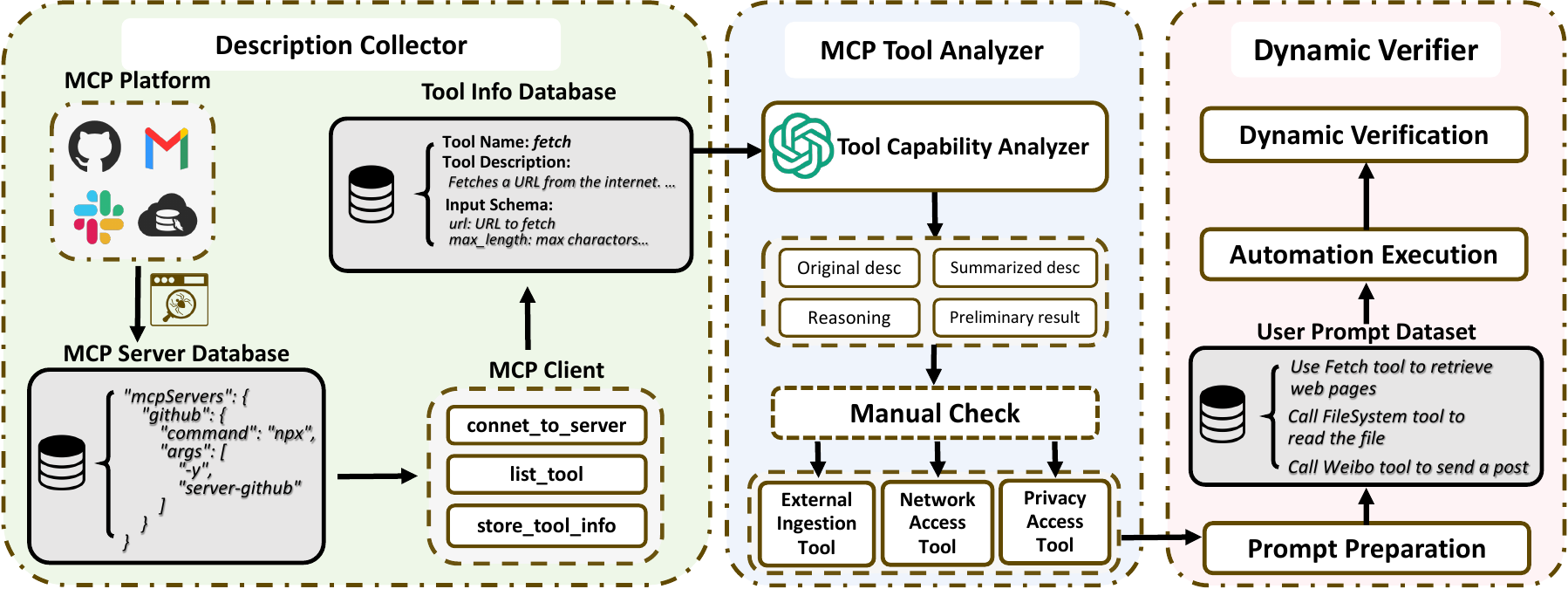}
    \caption{\textbf{The workflow of \ourtool.}}
    \label{fig:autotool}
\end{figure*}

To understand the impact of MCP-UPD at scale, it is not sufficient to demonstrate a single motivating case. What is needed is a systematic method to analyze the MCP ecosystem and identify the tools that can serve as building blocks for \textit{Parasitic Toolchain Attack}. Specifically, our goal is to measure the capability boundaries of MCP tools provided by publicly accessible MCP servers. We further determine whether they can be exploited as "gadgets" in the stages of MCP-UPD, namely, \textit{Parasitic Ingestion}, \textit{Privacy Collection}, and \textit{Privacy Disclosure}.

We design \ourtool, an analysis framework that can quantitatively assess risks across the MCP ecosystem: whether MCP servers expose tools that can be chained into full attack workflows. As illustrated in \autoref{fig:autotool}, the framework consists of three major components: a \textit{Description Collector} that crawls public MCP servers and extracts descriptions of tools from them, a \textit{MCP Tool Analyzer} that maps MCP tools to risk-related capabilities, and a \textit{Dynamic Verifier} that performs dynamic verification of tools with risk-related capabilities in the real-world environment. After passing through the dynamic verifier, the MCP tool can be proven capable of performing the risky operation in real-world environments.

\subsection{Description Collector}

The first stage of our framework focuses on discovering and collecting data about MCP servers and their exposed tools. Since MCP servers are often hosted on public platforms and open-source repositories, it is possible to systematically crawl and index them. This module aims to provide comprehensive coverage of the MCP ecosystem by collecting MCP server information and extracting tool descriptions as the foundation for further analysis.

\subsubsection{MCP Server Crawler}

First of all, it is necessary to collect comprehensive information about MCP servers. Since MCP servers hosted on public platforms typically provide corresponding GitHub links to make available as open-source projects for user download, we implemented the MCP Server Crawler to leverage these links as the entry point for data collection. The crawler gathers the names of MCP servers and their associated links of GitHub projects from hosting platform listings, then removes duplicates to ensure uniqueness. For each GitHub project, the crawler extracts detailed information, including the server name, execution commands, execution parameters, execution configurations, and repository statistics such as the number of GitHub stars. To facilitate MCP server deployment and ensure the feasibility of large-scale measurement analysis, the crawler collects only MCP servers packaged for direct execution via \texttt{npx}\cite{npx} or \texttt{uvx}\cite{uvx}, which encapsulate all necessary dependencies and configurations. The crawler then consolidates the collected information into the \textit{MCP Server Database}, which provides the structured information for subsequent automated analysis of tool capabilities.

\subsubsection{Tool Description Extractor}

After constructing the \textit{MCP Server Database}, the next step is to systematically extract information about all tools offered by these servers. Each MCP tool typically specifies three fields: name, description, and inputSchema, where the description explicitly defines the tool's function. Extracting this information enables us to analyze each tool's potential security risks.

To automate this extraction, we implemented an \textit{MCP Client} based on the official Python SDK~\cite{pythonsdk}, supporting connection establishment, session management, and tool invocation. The extraction steps are as follows: 
(1) The client connects to each MCP server using the server configuration commands in the database.
(2) The client issues a query to list all available tools provided by the MCP server. 
(3) The client retrieves the corresponding descriptions and stores the results in a structured \textit{Tool Information Database}.

The resulting \textit{Tool Information Database} contains a structured record of all discovered MCP tools across the analyzed MCP servers. For each tool, the database stores its name, description, and inputSchema, which serves as the foundation for subsequent analysis. 

\subsection{MCP Tool Analyzer}

With the tool descriptions collected, the second stage of our framework focuses on assessing the functional boundaries of tools and identifying which tools can be used for MCP-UPD. The rationale behind this stage is that MCP-UPD fundamentally relies on the inherent capabilities of tools. While some tools expose no meaningful attack surface and are therefore harmless, others provide functionalities that can be co-opted as "gadgets" in a parasitic toolchain, enabling adversaries to access or exfiltrate sensitive data. Therefore, systematically evaluating tool capabilities is essential to quantify the potential attack surface of MCP-UPD.

\subsubsection{Tool Capability Definition}

We define three capabilities of tools that are directly relevant to the MCP-UPD, as each phase of the attack chain depends on a specific type of functionality: \textit{External Ingestion Tools} enable \textit{Parasitic Ingestion}, \textit{Privacy Access Tools} support \textit{Privacy Collection}, and \textit{Network Access Tools} facilitate \textit{Privacy Disclosure}:

\squishlist

\item \textit{External Ingestion Tool}: Tools that can fetch textual data from external, potentially attacker-controlled sources (e.g., fetch, search, or social media connectors). These tools provide the initial entry point for parasitic prompts.

\item \textit{Privacy Access Tool}: Tools that can read sensitive local or user-specific data (e.g., read\_file, list\_contacts, search\_history). These tools act as the collection phase in the attack chain.

\item \textit{Network Access Tool}: Tools that can send or publish arbitrary text to external endpoints (e.g., send\_mail, postToWeb). These tools enable the disclosure phase by exfiltrating collected data.

\squishend

This taxonomy reflects the stages of the parasitic toolchain and allows us to assess whether a given MCP server contains the necessary "gadgets" of MCP-UPD.

\subsubsection{Tool Capability Analyzer}

To identify the target tools relevant to MCP-UPD, we leverage the semantic reasoning capabilities of LLMs. For each analysis, we provide the complete set of tools exposed by an MCP server as a unified input to the model, regardless of whether the server includes a single tool or many, enabling LLMs to reason over the tools collectively and identify those corresponding to predefined risk-related capabilities, including external ingestion, privacy access, and network disclosure, within the MCP server's natural deployment context. At the same time, the process balances efficiency with LLM token-length constraints, ensuring that the identification process remains scalable and accurate across the diverse categories of MCP servers.

We design prompt templates that instruct LLMs to interpret each tool's description and input schema, reason about its intended behavior, and map it onto one or more of the risk-related tool capabilities. For example, if a tool takes a URL parameter and returns arbitrary text, the model is prompted to consider whether this tool is an \textit{External Ingestion Tool} (e.g., description: "Fetches a URL from the internet", input parameter: url). Similarly, if a tool can both read local files and optionally send their contents to a remote endpoint, the model is prompted to consider its dual role as both a \textit{Privacy Access Tool} and a \textit{Network Access Tool} (e.g., description: "Reads the contents of a file and uploads them", input parameter: filepath), reflecting its potential to span multiple phases of the MCP-UPD attack chain. For each tool, the LLM outputs a quadruple consisting of its original description, the summarized description, the reasoning process, and the preliminary analysis result (i.e., the tool's capabilities). This quadruple serves as the basis for the subsequent manual check.

Because the analysis process based on LLMs can be affected by factors such as hallucination and inherent randomness, it is necessary to perform a manual check on the preliminary analysis results obtained by LLMs to ensure accuracy before conducting subsequent dynamic verification. In the manual check stage, we derive the manual check result based on the LLM-generated summarized description of the tool, the corresponding reasoning process, and the preliminary analysis result, while using the tool's original description as a reference. Three independent reviewers annotated tool capabilities. For each tool, disagreements were resolved via majority voting, and the resulting labels were treated as the ground truth for subsequent analysis.

\subsection{Dynamic Verifier}

The above tool capability analysis relies solely on information such as the tool descriptions provided in the MCP server, representing a static analysis approach. However, to actually trigger an MCP-UPD attack in real-world conditions, the tool must also be capable of executing its intended functionality in real-world environments. Therefore, a dynamic verification process is required to assess the tool’s real-world behavior and effectiveness, ensuring a rigorous determination of its practical risk-related capabilities.

The dynamic verifier can automatically perform validation based on real-world MCP Hosts (e.g., Cursor). The module consists of three stages: a user prompt dataset preparation stage, a batch automation execution stage, and a dynamic result analysis stage. 
During the user prompt dataset preparation, we generate user prompts corresponding to each tool invocation based on the tool capability analysis results above. For example, a tool with one risk-related capability generates one prompt, while a tool with two such capabilities generates two prompts. Each prompt explicitly specifies the tool name to ensure that the MCP host can correctly invoke the target tool for execution. In addition, we employ canary-style instrumentation, including controlled external inputs, synthetic sensitive artifacts, and explicit forwarding markers, to provide clear and verifiable detection signals. However, due to the diversity of MCP tool functionalities, some tools interact with external or environment-dependent resources beyond our experimental control. As a result, canary-based instrumentation is not universally applicable across all tools.

In the batch automation execution stage, we employ a keyboard–mouse simulator to perform automated interactions, automatically sending the generated user prompt to the MCP host, and then awaiting the returned results. Meanwhile, the tool execution adopts a linear switching mechanism: after one tool finishes execution, the next tool will be executed automatically. After the entire process, it outputs complete chat logs for all tools under the given user prompts, including user-sent messages, tool invocation traces, and the corresponding agent responses.

During the dynamic result analysis, given the large number of tools under evaluation and the complexity of the collected data, we first employ LLMs for preliminary analysis. Each agent chat log containing tool invocation traces is fed into the LLM, which determines whether the tool exhibits a specific capability and provides the corresponding reasoning behind the judgment. For each capability, the verification follows the rules below:

\squishlist

\item \textit{External Ingestion Tool}: 
Chat logs are analyzed to determine whether controlled external inputs or other uncontrolled target content are retrieved and incorporated into the conversation context, indicating successful ingestion.

\item \textit{Privacy Access Tool}: 
Chat logs are analyzed to determine whether the tool accesses synthetic sensitive artifacts or other targeted sensitive data (e.g., confidential file contents, environment variables, or application message histories).

\item \textit{Network Access Tool}: 
Chat logs are analyzed to determine whether a network-related MCP tool is invoked with parameters containing synthetic sensitive artifacts or other targeted sensitive data, indicating potential data exfiltration.

\squishend

To ensure the reliability of dynamic verification, we perform a manual check. Specifically, three independent reviewers manually inspect the LLM's reasoning, cross-reference it with the corresponding agent chat logs, and examine the invocation parameters and returned results. Disagreements are resolved via majority voting, and the resulting labels are treated as the final outputs of the dynamic verifier.
Finally, after the dynamic verification, we can identify each tool's risk-related capabilities within practical environments.

\begin{table}[t]

\centering
\caption{The sources of MCP servers and MCP tools.}
\label{tab:sources}
\resizebox{0.45\textwidth}{!}{
\begin{tabular}{ l c c }
\toprule
\textbf{Sources} & \textbf{MCP Server} & \textbf{MCP Tool} \\

\midrule
    Pulse MCP\cite{pulsemcp} & 784 (57.6\%) & 6,736 (55.1\%) \\    
    MCP Market\cite{mcpmarket}  & 310 (22.8\%) & 2,781 (22.8\%) \\
    Awesome MCP Servers\cite{awesome} & 266 (19.6\%) & 2,713 (22.2\%) \\
\midrule
\textbf{Total} & \textbf{1,360} & \textbf{12,230} \\
\bottomrule
\end{tabular}
}
\end{table}

\section{Evaluation}
\label{evaluate}

\begin{table}[t]
\centering
\caption{Category distribution of MCP servers.}

\label{tab:category_server}
\setlength{\tabcolsep}{6pt}
\resizebox{0.42\textwidth}{!}{
\begin{tabular}{l c c }
\toprule
\textbf{Category} & \textbf{Number} & \textbf{Proportion} \\
\midrule
Information Retrieval & 152 & 11.2\%  \\
AI-Driven Utilities & 146 & 10.7\%  \\
Project Mgmt. \& Collaboration & 125 & 9.2\% \\
Development \& Testing & 110 & 8.1\%  \\
Data Processing \& Analytics & 109 & 8.0\%  \\
Docs. \& Knowledge Bases & 101 & 7.4\%  \\
Blockchain \& Financial Systems & 85 & 6.3\% \\
Multimedia Processing & 84 & 6.2\% \\
Command Execution & 71 & 5.2\% \\
Cloud Services & 62 & 4.6\% \\
Security \& Monitoring & 57 & 4.2\% \\
Communication \& Email & 54 & 4.0\%  \\
Social Media Platforms & 36 & 2.6\% \\
Geospatial \& Transportation & 28 & 2.1\% \\
\midrule
Others    & 140 & 10.3\% \\
\midrule
Total & 1,360 & - \\
\bottomrule
\end{tabular}
}
\end{table}

In this section, we discuss the analysis results of \ourtool and focus on three research questions.

\paragraph{Experiment Setup} \ourtool runs in a real home environment on a MacBook with macOS 15.6.1 and 16 GB of memory.

\paragraph{Dataset} 
In this paper, we collected MCP servers from three sources, including two public hosting platforms (Pulse MCP\cite{pulsemcp} and MCP Market\cite{mcpmarket}) and one open-source repository (Awesome MCP Servers\cite{awesome}).
To facilitate MCP server deployment and ensure the feasibility of large-scale measurement analysis, we focused on MCP servers packaged for direct execution with \texttt{npx}\cite{npx} or \texttt{uvx}\cite{uvx}, which bundle all required dependencies and configurations.
Finally, we collected 1,360 MCP servers. These servers exposed 12,230 tools, which formed the dataset for our subsequent measurement and analysis. The sources of these servers and tools are detailed in \autoref{tab:sources}.

To support our analysis, we examined not only the number of MCP servers but also their functional categories, as different contexts entail distinct risks (i.e., ingestion via information retrieval or social media, privacy disclosure via command execution or cloud services). We used LLMs to semantically parse server descriptions and documentation, classifying them into high-level categories, such as Information Retrieval, Command Execution, and Communication \& Email, with ambiguous cases cross-validated across multiple models. \autoref{tab:category_server} presents the category distribution of our collected MCP servers, with the Others category combining numerous small categories with a few servers.

\begin{table*}[t]
\centering
\small
\caption{Cases of manual checks in tool capability analysis.}
\label{tab:manual_check_examples}
\begin{tabular}{p{4.5cm}p{12.5cm}}
\toprule
\textbf{Check Items} & \textbf{Content} \\
\midrule
\textbf{Tool Name} (MCP Server Name) & \texttt{read\_file} (Filesystem) \\
Tool Description & Reads the entire content of a specified file as UTF-8 text. \\
Summary Description & Reads local file contents. \\
Reasoning & The tool reads file contents from a specified path, indicating private data access. \\
LLM Result & Privacy Access Tool (PAT). \\
Manual Check & $\Box$ (\ding{51} or \ding{55}) 
\\
\midrule
\textbf{Tool Name} (MCP Server Name) & \texttt{unified\_search} (Unified Search) \\
Tool Description & Search across multiple online sources, such as Google Scholar, Google Web Search, YouTube, through a unified interface and return relevant results for the query. \\
Summary Description & Unified search over multiple connected data sources. \\
Reasoning & The tool searches across multiple connected sources, suggesting that it can retrieve a broad range of information. \\
LLM Result & External Ingestion Tool (EIT). \\
Manual Check & $\Box$ (\ding{51} or \ding{55})  \\
\bottomrule
\end{tabular}
\end{table*}

\paragraph{Analysis Efficiency} 
For all 1,360 servers and 12,230 tools, \ourtool completed 255 hours of automated analysis, averaging 675 seconds per server and 75 seconds per tool. Manual checks required 71 hours (189 seconds per server and 21 seconds per tool) for tool capability analysis and 143 hours (379 seconds per server and 42 seconds per tool) for dynamic verification. The efficiency of manual checks varies significantly with tool complexity. Most tools have simple and well-defined functionalities (e.g., \texttt{read\_file}), allowing reviewers to verify LLM-generated outputs against tool specifications within about 10 seconds. In contrast, more complex tools (e.g., \texttt{unified\_search}), which operate over heterogeneous data sources, require assessing their access scope and typically take 50–60 seconds to verify. Manual effort per tool in dynamic verification is higher than in tool capability analysis, as simpler tools are filtered out in earlier stages, and the remaining cases involve more complex execution traces, requiring additional inspection of invocation parameters and returned results. Overall, the results indicate that manual verification remains practical and scalable, given the predominance of simple tools and the staged filtering design.

\autoref{tab:manual_check_examples} illustrates two representative cases (\texttt{read\_file} and \texttt{unified\_search}) from the manual check process. Each case includes the tool name, the related MCP Server name, the original tool description, the LLM-generated summary, the model's reasoning, and its predicted result. Reviewers examine this information to determine whether the prediction is correct and document their decision (correct or incorrect) in the annotation box.

\paragraph{Research Questions} To systematically evaluate the security risks posed by MCP-UPD, we design our large-scale measurement around three key research questions: 
\squishlist
    \item \textbf{RQ1.} \textbf{How prevalent are MCP tools in the MCP ecosystem that can be exploited as "gadgets" in the MCP-UPD?} Understanding the prevalence of risky tools is essential to assess the baseline attack surface of the MCP ecosystem. By quantifying how many tools provide capabilities relevant to MCP-UPD, we can determine whether such tools are isolated anomalies or a widespread phenomenon that attackers can readily exploit.
    \item \textbf{RQ2.} \textbf{How many MCP servers expose tools that could be exploited as building blocks of MCP-UPD toolchains?} Shifting the perspective from tools to servers provides insight into the real deployment risks of the MCP ecosystem. By analyzing how many servers aggregate exploitable tools, we can assess the likelihood of attackers encountering viable entry points and identify high-risk servers that concentrate multiple dangerous capabilities.
    \item \textbf{RQ3.} \textbf{How effective are MCP-UPD toolchains in real-world scenarios?} To further evaluate the real-world threat of MCP-UPD, we need to construct toolchains based on exploitable "gadgets" and evaluate their effectiveness to align with real-world attack scenarios. Besides, we need to evaluate the performance of MCP-UPD toolchains across different LLMs and clients.
\squishend

\begin{table}[t]
\centering
\caption{LLM analyzing accuracy in tool capability analysis and dynamic verification.}
\label{tab:llm_accuracy}
\resizebox{0.49\textwidth}{!}{
\begin{tabular}{lcccc}
\toprule
\multirow{2}{*}{Category} & \multicolumn{2}{c}{Tool Capability Analysis} & \multicolumn{2}{c}{Dynamic Verification} \\
\cmidrule(lr){2-3} \cmidrule(lr){4-5}
 & Precision (\%) & Recall (\%) & Precision (\%) & Recall (\%) \\
\midrule
EIT & 86.5 & 99.2 & 95.1 & 99.2 \\
PAT & 85.4 & 95.8 & 94.5 & 98.2 \\
NAT & 81.1 & 99.6 & 96.4 & 98.9 \\
\midrule
Overall & 85.5 & 98.0 & 95.1 & 98.8 \\
\bottomrule
\end{tabular}
}
\end{table}

\subsection{Statistics of Exploitable MCP Tools}
\label{subsec:mcptools}

\paragraph{Overall result} \textit{Among all 12,230 tools collected from MCP servers, the tool capability analysis identified 5,666 tools as potentially risky. Following dynamic verification, 1,062 tools (8.7\% of the total) were confirmed to possess at least one risk-related capability exploitable in MCP-UPD toolchains.}

This result shows that exploitable capabilities are not rare exceptions, but are widely present across the MCP ecosystem. Importantly, this subset represents the \textbf{operational attack surface} of MCP, that is, the tools that an attacker can realistically leverage to construct MCP-UPD toolchains.
As a result, attackers do not need to rely on specially crafted or rare tools, but can easily find and utilize exploitable tools, making MCP-UPD both practical and broadly applicable in real-world deployments.
 
\begin{figure}[t]
    \centering
    \includegraphics[width=0.48\textwidth]{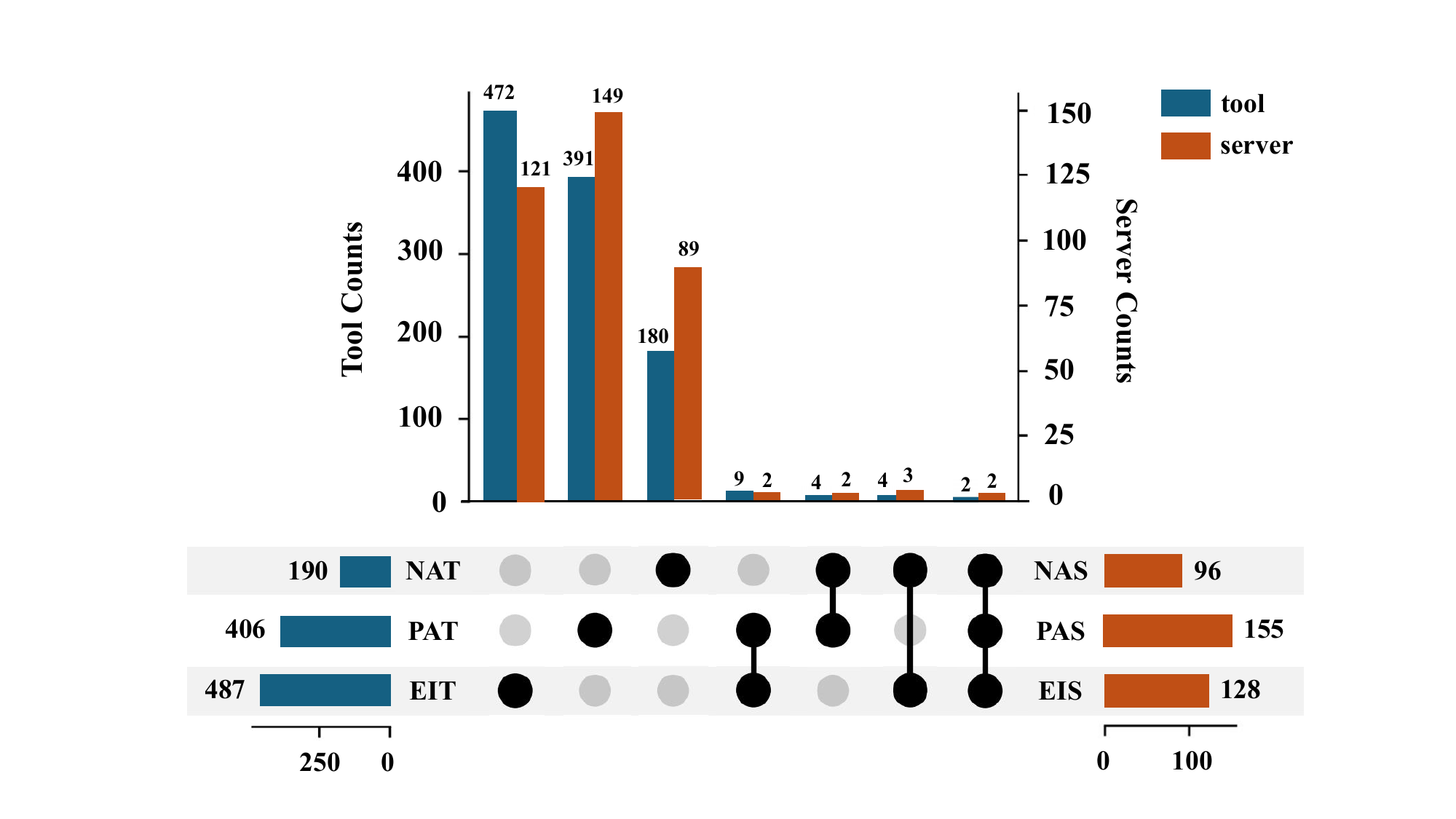}
    \caption{\textbf{Statistics of exploitable MCP tools/servers}. EIT/S represents the External Ingestion Tool/Server, PAT/S represents the Privacy Access Tool/Server, NAT/S represents the Network Access Tool/Server.}
    \label{fig:tool_statistics}
\end{figure}

After further analyzing these 1,062 tools, we revealed the detailed distribution of their risk-related capabilities, as shown in \autoref{fig:tool_statistics}. Specifically, 472 tools were identified solely as External Ingestion Tools (EIT), 391 tools solely as Privacy Access Tools (PAT), and 180 tools solely as Network Access Tools (NAT). In addition, 9 tools qualified for both EIT and PAT, 4 tools for both PAT and NAT, 4 tools for both EIT and NAT, and 2 tools for all three types of capabilities. In total, 487 tools were identified as EIT, 406 as PAT, and 190 as NAT. 
The above data reveals that at each key step of the MCP-UPD toolchain, there exists a nontrivial set of MCP tools that can be practically used in real-world attacks. Some tools can complete multiple steps and even fully cover the entire MCP-UPD attack.

\begin{table*}[t]
\centering
\caption{High-risk MCP servers and their detailed risk distribution.}
\label{tab:server_risks}

\scriptsize \textbf{IR}: Information Retrieval, \textbf{PMC}: Project Management \& Collaboration,  \textbf{SMP}: Social Media Platforms, \textbf{ADU}: AI-Driven Utilities \textbf{DT}: Development \& Testing \textbf{DKB}: Document \& Knowledge Bases, \textbf{CEx}: Communication \& Email \textbf{Risk P.}: Risk Tool's Proportion, \textbf{EIT C. / P.}: External Ingestion Tool Count / Proportion, \textbf{PAT C. / P.}: Privacy Access Tool Count / Proportion, \textbf{NAT C. / P.}:  Network Access Tool Count / Proportion .

\setlength{\tabcolsep}{5pt}
\resizebox{1\textwidth}{!}{
\begin{tabular}{l l l c c c c c c c c c c }
\toprule
\textbf{Server Name}  & \textbf{Category} & \textbf{Source} & 
\textbf{Tools} & \textbf{Risk Tools} & \textbf{Risk P.} & 
\textbf{EIT C.} & \textbf{EIT P.} & 
\textbf{PAT C.} & \textbf{PAT P.} & 
\textbf{NAT C.} & \textbf{NAT P.} & 
\textbf{Risks} \\
\midrule
Bright Data\cite{brightdata}                 & IR & PulseMCP\cite{pulsemcp}   & 65 & 18 & 27.7\% & 18  & 27.7\%  & 0 & 0.0\% & 0 & 0.0\% & 18 \\
Gitee\cite{gitee}                & PMC & PulseMCP\cite{pulsemcp}   & 20 & 18 & 90.0\% & 18 & 90.0\% & 0  & 0.0\%  & 0 & 0.0\% & 18  \\
Chinese Trends Hub\cite{cth}                  & SMP & PulseMCP\cite{pulsemcp}   & 21 & 17 & 81.0\% & 17 & 81.0\% & 0  & 0.0\%  & 0 & 0.0\% & 17  \\
Discogs\cite{discogs}               & IR          & AwesomeMCP\cite{awesome} & 42 & 15 & 38.1\% & 15 & 35.7\% & 0  & 0.0\%  & 1  & 2.4\%  & 16  \\
Agentic Tools\cite{agentic}                      & ADU          & MCP Market\cite{mcpmarket}   & 20 & 15 & 75.0\% & 0 & 0.0\% & 15  & 75.0\%  & 0  & 0.0\%  & 15  \\
MCP Connect\cite{mcpconnect} & DT& MCP Market\cite{mcpmarket}   & 26 & 13 &59.1\% & 13 &50.0\% & 0  & 0.0\%  & 0  & 0.0\%  & 13  \\
Contentful Management\cite{contentful}                   & DKB         & AwesomeMCP\cite{awesome}   & 31 & 11 &35.5\% & 11  & 35.5\%  & 0 & 0.0\% & 0  & 0.0\%  & 11  \\
DesktopCommanderMCP\cite{dcmcp}            & CEx   & AwesomeMCP\cite{awesome}  & 18 & 10 & 55.6\% & 0  & 0.0\%  & 18 & 55.6\% & 0  & 0.0\%  & 10  \\
Hacker News\cite{hnmcp}                    & SMP          & PulseMCP\cite{pulsemcp}  & 11 & 9 & 81.8\% & 0 & 0.0\% & 0  & 0.0\% & 9 & 81.8\% & 9  \\
HAL\cite{hal}                    & DT          & PulseMCP\cite{pulsemcp}   & 8 & 8 & 100.0\% & 6 & 75.0\% & 0  & 0.0\%  & 2  & 25.0\%  & 8  \\
\bottomrule
\end{tabular}}
\end{table*}

In particular, two MCP tools can provide all three capabilities. While capable of executing MCP-UPD independently, their versatility also allows them to function at any stage and combine seamlessly with single-capability tools in MCP-UPD toolchains, greatly broadening options for attacker flexibility.
For example, with \texttt{execute-command}, an attacker could use it in any stage in MCP-UPD:

\squishlist

\item \textbf{EIT}. The attacker embeds malicious prompts in a public resource (e.g., an online document). When the victim uses \texttt{execute-command} to fetch it (e.g., \texttt{curl http://public.com/contents}), the injected prompt can trigger other PATs and NATs for malicious use. 

\item \textbf{PAT}. The malicious prompt ingested by other EITs guides the LLM to use \texttt{execute-command} for sensitive data access (e.g., \texttt{cat mcp.json}), such as reading local configuration files that store API keys.

\item \textbf{NAT}. The malicious prompt ingested by other EITs guides the LLM to send the privacy information obtained by other PATs to a target receiver using \texttt{execute-command} (e.g., \texttt{curl -X POST -F file=sk-xxx http://attacker.com/upload}), enabling the attacker to receive the victim's privacy data.

\squishend

Besides, from the perspective of overlaps and combinations, although most tools are single-purpose, the distributions of EIT, PAT, and NAT are highly complementary. Within the MCP-UPD attack paradigm of multi-tool tasks, this complementarity enables attackers to construct complete MCP-UPD toolchains within or across MCP servers in real-world scenarios.

To evaluate the reliability of our LLM-based analysis and mitigate subjective bias, we measure its precision and recall with respect to manual annotations, which are treated as ground truth. As shown in \autoref{tab:llm_accuracy}, the LLM achieves a precision of 85.5\% and a recall of 98.0\% in the tool capability analysis stage. In the dynamic verification stage, the precision is 95.1\%, and the recall is 98.8\%. These results indicate that the LLM achieves high recall while maintaining reasonable precision, thereby reducing the manual verification effort.
Furthermore, we assess inter-annotator agreement among the three reviewers using Fleiss' Kappa, obtaining a score of 0.72, which indicates substantial agreement.

\begin{mybox}[boxsep=0pt,
	boxrule=1pt,
	left=4pt,
	right=4pt,
	top=4pt,
	bottom=4pt,
	]
~\textbf{Answer to RQ1:} 
8.7\% of MCP tools expose risk-related capabilities, and although most cover only a single stage, the complementarity of three capabilities enables complete MCP-UPD toolchains. Two tools even provide all three capabilities, allowing them to combine flexibly with other tools at any stage of MCP-UPD.
\end{mybox}

\subsection{MCP Servers with Risky Tools}
\label{subsec:servers_tools}

\paragraph{Overall result} \textit{Among all 1,360 MCP servers we collected, 370 servers contain at least one threat-relevant MCP tools, accounting for 27.2\% of the total servers.} 

When shifting the analysis from tools to servers, we observe that a nontrivial number of MCP servers expose tools that can form MCP-UPD attack toolchains, as shown in \autoref{fig:tool_statistics}. Specifically, 121 servers contain at least one External Ingestion Tool (EIT), 149 servers include a Privacy Access Tool (PAT), and 89 servers have a Network Access Tool (NAT). Overlaps can further amplify risks: 2 servers host both EIT and PAT, 3 servers host PAT and NAT, and 2 servers host EIT and NAT. Notably, 2 servers support all three tool capabilities, enabling independent execution of MCP-UPD. Overall, servers having EIT are the most prevalent (128), followed by PAT (155) and NAT (96).
This distribution reveals two key insights. First, although most MCP servers offer only single-capability tools, nearly 30\% provide exploitable components that can be dynamically invoked in real-world scenarios, showing that risky servers form a nontrivial portion of the ecosystem rather than being limited to a few isolated cases. Second, two servers aggregate EIT, PAT, and NAT, enabling complete MCP-UPD toolchains within a single server without cross-server coordination. These findings demonstrate that MCP-UPD is practical using common servers with complementary capabilities.

We further analyze the 10 MCP servers hosting the most risky tools, as detailed in \autoref{tab:server_risks}. The number of risky tools ranges from 8 in \texttt{HAL} to 18 in \texttt{Bright Data}. \texttt{HAL} have a 100\% risky tool ratio, while \texttt{Bright Data} shows the lowest (27.7\%). Among tool capabilities, EITs are the most common, appearing in 7 servers, with \texttt{Bright Data} and \texttt{Gitee} hosting the most (18). PATs are present in 2 servers, with \texttt{Agentic Tools} having 15 and \texttt{DesktopCommanderMCP} having 18, while NATs appear in 3 servers, peaking at 9 in \texttt{Hacker News}.

Categorically, \texttt{Information Retrieval} (IR), \texttt{Social Media Platforms} (SMP), and \texttt{Development \& Testing} (DT) appear most frequently in the high-risk set (two servers each), followed by single instances from \texttt{Project Management \& Collaboration} (PMC), \texttt{AI-Driven Utilities} (ADU), \texttt{Documents \& Knowledge Bases} (DKB), and \texttt{Command Execution} (CEx). Furthermore, clear correlations appear between server functionality and attack capability. Information retrieval and aggregation platforms (\texttt{Bright Data}, \texttt{Discogs}) tend to expose EIT capabilities, as their core functionality relies on ingesting external data through retrieval interfaces. Social media platforms (\texttt{Chinese Trends Hub}, \texttt{Hacker News}) exhibit very high densities of ingestion tools that can be repurposed as external entry points into the toolchain, and in some cases also expose network-access capabilities that enable data exfiltration. In contrast, workflow and agent-oriented platforms such as \texttt{Agentic Tools} and \texttt{DesktopCommanderMCP} emphasize command execution and cross-tool combination, naturally aligning with the privacy access stages of MCP-UPD. Development-oriented servers (\texttt{MCP Connect}, \texttt{HAL}) and management-focused platforms (\texttt{Contentful Management}, \texttt{Gitee}) concentrate heavily on ingestion-oriented functionality, as their core purpose involves synchronizing resources. Notably, both \texttt{HAL} and \texttt{Discogs} expose two distinct MCP-UPD–relevant capabilities within the same server, making them particularly easy to repurpose into flexible toolchains. 

\begin{table*}[t]
\centering
\caption{Distribution of attack capabilities across MCP server functional categories.}

\label{tab:category_risk}
\scriptsize
\textbf{Bold numbers} indicate the highest values in each attack stage.
\setlength{\tabcolsep}{6pt}
\resizebox{0.98\textwidth}{!}{
\begin{tabular}{l r r r r r r}
\toprule
\multirow{2}{*}{\textbf{Category}} & 
\multicolumn{2}{c}{\textbf{Parasitic Ingestion}} &
\multicolumn{2}{c}{\textbf{Privacy Collection}} &
\multicolumn{2}{c}{\textbf{Privacy Disclosure}} \\
\cmidrule(lr){2-3}\cmidrule(lr){4-5}\cmidrule(lr){6-7}
 & Counts(Rank) & Proportions(Rank) &  Counts(Rank) & Proportions(Rank) & Counts(Rank) & Proportions(Rank) \\
\midrule
Information Retrieval  & \textbf{36 (1)} & 23.7\% (2) & 4 (12) & 2.6\% (12) & 5 (6) & 3.3\% (10) \\
AI-Driven Utilities & 8 (7) & 5.5\% (11) & 19 (2) & 13.0\% (5) & 3 (9) & 2.1\% (12) \\
Project Management \& Collaboration & 7 (8) & 5.6\% (9) & 5 (10) & 4.0\% (11) & 19 (2) & 15.2\% (2) \\
Development \& Testing  & 10 (3) & 9.1\% (5) & \textbf{32 (1)} & \textbf{29.1\% (1)} & 4 (7) & 3.6\% (9) \\
Data Processing \& Analytics & 2 (11) & 1.8\% (12) & 13 (4) & 11.9\% (6) & 3 (9) & 2.8\% (11) \\
Documents \& Knowledge Bases & 9 (6) & 8.9\% (6) & 11 (6) & 10.9\% (8) & 6 (4) & 5.9\% (5) \\
Blockchain \& Financial Systems & 13 (2) & 15.3\% (3) & 6 (9) & 7.1\% (10) & 11 (3) & 12.9\% (3) \\
Multimedia Processing  & 1 (13) & 1.2\% (14) & 10 (7) & 11.9\% (6) & 0 (13) & 0.0\% (13) \\
Command Execution & 10 (3) & 14.1\% (4) & 18 (3) & 25.4\% (2) & 4 (7) & 5.6\% (6) \\
Cloud Services  & 1 (13) & 1.6\% (13) & 12 (5) & 19.4\% (3) & 6 (4) & 9.7\% (4) \\
Security \& Monitoring  & 4 (9) & 7.0\% (8) & 5 (10) & 8.8\% (9) & 3 (9) & 5.3\% (8) \\
Communication \& Email & 3 (10) & 5.6\% (9) & 9 (8) & 16.7\% (4) & \textbf{20 (1)} & \textbf{37.0\% (1)} \\
Social Media Platforms & 10 (3) & \textbf{27.8\% (1)} & 0 (13) & 0.0\% (13) & 2 (12) & 5.6\% (6) \\
Geospatial \& Transportation  & 2 (11) & 7.1\% (7) & 0 (13) & 0.0\% (13) & 0 (13) & 0.0\% (13) \\

\bottomrule
\end{tabular}
}
\end{table*}

\begin{figure}[t]
    \centering
    \includegraphics[width=0.46\textwidth]{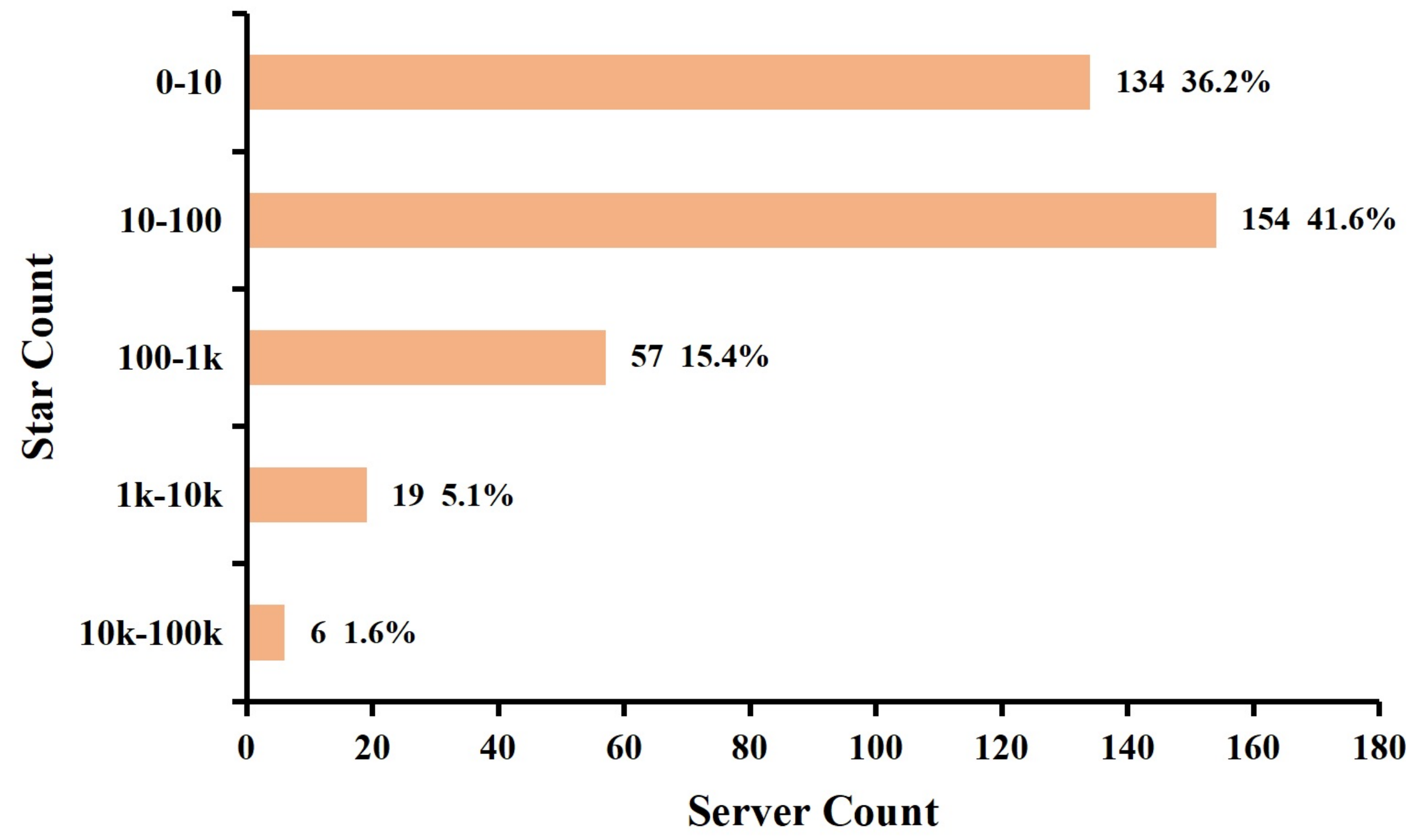}
    \caption{\textbf{Distribution of GitHub stars of MCP servers with risky tools}.}
    \label{fig:star_statistics}
\end{figure}

Moreover, we assessed the popularity of MCP servers with risky tools via their GitHub stars. As shown in \autoref{fig:star_statistics}, 82 servers (22\%) have 100+ stars, indicating that many projects with risky tools gained notable developer attention. Furthermore, several servers exceed 1,000 stars, suggesting that even widely adopted MCP servers face significant tool exposure risks. Thus, these security issues extend beyond obscure projects to popular ones as well. Although most servers remain in the low-star range (0–100), this is expected, as MCP is new and many servers were only recently released, without sufficient time to build community visibility, which does not reflect a lack of community interest.

As discussed above, the attack capability of an MCP server is related to its function type. \autoref{tab:category_risk} shows the distribution of attack capabilities across the functional categories of MCP servers. Almost all MCP server categories 90.5\%) can cover all three attack capabilities. 
Only \texttt{Geospatial \& Transportation} and \texttt{Multimedia Processing} lacks the \textit{Privacy Disclosure} capability, as their primary functionality focuses on rendering or transforming content rather than transmitting user data to external endpoints.
For \textit{Parasitic Ingestion}, \texttt{Information Retrieval} servers rank highest in count and \texttt{Social Media Platforms} rank highest in proportion, both serving as primary entry points for parasitic ingestion. They focus on aggregating large volumes of external content inherently expand ingestion surfaces, making them especially prone to malicious prompt injection. 

For \textit{Privacy Collection}, \texttt{Development \& Testing} servers rank highest in both count and proportion. These platforms frequently read configuration files, logs, and project artifacts that may contain credentials or sensitive personal data, and their deep integration with organizational workflows makes such information particularly susceptible to harvesting when misused in toolchain.
For \textit{Privacy Disclosure}, \texttt{Communication \& Email} servers rank highest in both count and proportion. These platforms are inherently designed to transmit messages and content to external recipients, and their tools routinely provide outward communication capabilities. When misused in MCP-UPD, these outward-facing channels can be easily repurposed to exfiltrate sensitive information to the public Internet, leading to direct privacy disclosure.
Beyond the top-ranked categories, others also play key roles in MCP-UPD. For \textit{Parasitic Ingestion}, \texttt{Project Management \& Collaboration} servers can expose shared tasks, documents, and integrations, For \textit{Privacy Collection} and \textit{Disclosure}, \texttt{Command Execution} servers also contribute significantly, as their flexible execution interfaces can both access sensitive data and transmit it to external endpoints, whereas \texttt{Cloud Services} centralize multi-tenant data, having both collection and disclosure impacts.
This functional "division of labor" enables MCP-UPD toolchains to be assembled in diverse, adaptive ways, depending on available servers.

\begin{mybox}[boxsep=0pt,
	boxrule=1pt,
	left=4pt,
	right=4pt,
	top=4pt,
	bottom=4pt,
	]
~\textbf{Answer to RQ2:} 
27.2\% of MCP servers expose at least one exploitable tool, with many widely used servers that amplify MCP-UPD exposure. Moreover, nearly all server categories contain the three attack capabilities, which enable adaptive and diverse attack compositions across the ecosystem.
\end{mybox}

\begin{table*}[t]
\centering
\caption{Attack success rate across evaluated toolchains.}
\label{tab:toolchain_results}

\scriptsize \textbf{Popular}: Github Stars \& Server Downloads

\setlength{\tabcolsep}{4pt}
\resizebox{1\textwidth}{!}{
\begin{tabular}{l l l r l l r l l r c}
\toprule
\multirow{2}{*}{\textbf{ID}} &
\multicolumn{3}{c}{\textbf{External Ingestion}} &
\multicolumn{3}{c}{\textbf{Privacy Access}} &
\multicolumn{3}{c}{\textbf{Network Access}} &
\multirow{2}{*}{\textbf{Success}} \\
\cmidrule(lr){2-4}\cmidrule(lr){5-7}\cmidrule(lr){8-10}
 & Server & Tool & Popular &
   Server & Tool & Popular &
   Server & Tool & Popular &
   Trials \\
\midrule
1  & fetch        & fetch                       & 72.5k \& 23.4m &
     FileSystem    & read\_file                  & 72.5k \& 78.7k &
     Gmail         & send\_mail                  & 814 \& 116k  & 8/10 \\
2  & GitMCP       & fetch\_generic\_url\_content & 6.9k \& 989k  &
     Everything    & printEnv                    & 72.5k \& 368k &
     Slack         & slack\_post\_message        & 72.5k \& 519k  & 2/10 \\
3  & Playwright   & playwright\_get             & 23.1k \& 23.5k  &
     DesktopCommander & get\_config              & 4.9k \& 702k &
     Notion        & API-patch-block-children   & 3.5k \& 434k  & 7/10 \\
4  & HAL         & http-get            & 34 \& 4.6k   &
     haris-musa/excel-mcp-server & read\_data\_from\_excel & 2.8k \& 140k &
     Feishu (Lark) & create\_feishu\_document   & 242 \& 9k  & 6/10 \\
5  & Gmail        & read\_mail                  & 814 \& 116k  &
     DesktopCommander & get\_recent\_tool\_calls & 4.9k \& 702k &
     Mcp-Proxy     & create\_issue               & 578 \& -  & 3/10 \\
6 & Web Fetcher  & fetch\_url                  & 908 \& 73.6k   &
     Slack         & slack\_get\_channel\_history & 72.5k \& 519k &
     instantly     & create\_campaign            & 13 \& 1.8k & 4/10 \\
7  & Web Research & visit\_page                 & 285 \& 22.8k  &
     SQL Alchemy   & execute\_query              & 360 \& 51k   &
     linear        & Linear\_createIssue         & 117 \& 16.1k  & 1/10 \\
8  & Bright Data  & scrape\_as\_html            & 1.6k \& 59.3k &
     Mondaycom API & get\_board\_items\_by\_name & 334 \& 47.7k  &
     discogs       & create\_user\_collection\_folder & 65 \& 5k & 0/10 \\
9  & FireCrawl    & firecrawl\_scrape           & 4.9k \& 691k &
     linear        & Linear\_getProjects         & 117 \& 16.1k  &
     Mondaycom API & create\_item                & 334  \& 47.7k  & 2/10 \\
10  & ms-365       & get-mail\_message           & 334 \& 46.8k  &
     ms-365        & list-mail-message           & 334 \& 46.8k  &
     ms-365        & send-mail                   & 334 \& 46.8k  & 3/10 \\
\bottomrule
\end{tabular}
}
\end{table*}

\subsection{MCP Toolchains Evaluation}

\paragraph{Overall result} \textit{MCP-UPD can be completed in 9 of 10 toolchains, 5 of 7 LLMs, and all five clients.}

\paragraph{MCP Toolchains Experiment}
In the above experiments, we identified a large number of risky tools within the MCP ecosystem. To further validate the real-world impact of MCP-UPD, we conduct a series of empirical studies by deploying representative MCP servers and observing how these tools behave in practical interaction scenarios.
Specifically, we sample tools having the three risk-related capabilities (external ingestion, privacy access, and network access) based on popularity metrics (i.e., GitHub stars and server downloads). The sampling prioritizes highly popular servers to better reflect real-world impacts while maintaining diversity across different server types to adapt to diverse usage scenarios within the MCP ecosystem.
For each risk-related capability, we select 10 representative tools. Then, we construct 10 tool invocation chains for real-world evaluation. Among them, nine chains are formed by randomly pairing tools across different servers, while one chain uses three tools from the same server to illustrate a self-contained attack chain achievable within a single MCP server.
The procedure of the MCP toolchains experiment is as follows:

\squishlist

\item \textbf{Preparation}. For the \textit{External Ingestion Tool}, we publish an application page or web page containing a crafted malicious prompt that guides the agent to invoke the \textit{Privacy Access Tool} to obtain sensitive data and the \textit{Network Access Tool} to exfiltrate it to the target receiver.

\item \textbf{Normal usage simulation}. First, we add the test MCP servers to Cursor's MCP configuration. Then, we simulate a victim's routine use in a real-world environment. Specifically, the user issues a routine request in Cursor, such as \textit{"Please fetch the content at www.example\_attack.com"}, triggering the \textit{External Ingestion Tool} to retrieve content from a malicious page containing the crafted prompt.

\item \textbf{Result verification}. After Cursor completes the operation, we determine whether the attack succeeded by checking if the privacy data was sent by the \textit{Network Access Tool} to the target receiver.

\squishend

Considering model randomness, we repeat each end-to-end toolchain experiment 10 times. A toolchain is labeled successful if any trial results in verified exfiltration. All experiments are conducted in Cursor using its default Cursor Auto model. To evaluate the technical exploitability of MCP-UPD, we minimize human-in-the-loop effects by configuring the Cursor client for automated execution.

\paragraph{Results}
Table \ref{tab:toolchain_results} shows the evaluation results of the 10 toolchains, which derive the following three key observations.

\squishlist

\item \textbf{Obs\#1}. The majority of constructed toolchains can be successfully exploited: 9 of 10 toolchains result in confirmed privacy exfiltration in at least one of 10 trials. We observe consistent injection of attacker-controlled external content into the LLM context across all repeated trials for each evaluated toolchain. These results suggest that MCP-UPD is not an isolated phenomenon but a repeatable threat across diverse real-world MCP servers.

\item \textbf{Obs\#2}. Three toolchains (Chains 1, 3, and 4) exhibit successful exfiltration in 6–8 out of 10 trials. It shows that once an MCP-UPD toolchain is reachable, attackers can consistently exploit the toolchain's attack trajectory, even in the face of LLMs' inherent randomness, without adversarial tuning or repeated prompt engineering.

\item \textbf{Obs\#3}. Chain 10 uses three tools from the same server, each providing a distinct capability. When the ingestion tool retrieves an e-mail containing the parasitic prompt, it triggers a chain of internal calls, listing and then sending historical e-mails, showing that a single server with all three capabilities can inherently execute a complete MCP-UPD attack.

\item \textbf{Obs\#4}. Chain 8 fails in all 10 trials. Further analysis revealed significant differences in functionality among the three tools (crawling web content, obtaining work plans, and publishing music collection names), making it difficult to induce the LLM to invoke these tools naturally. By combining with other successful toolchains, we can find that the functional relevance among tools affects MCP-UPD success: stronger relevance yields higher success rates, while weaker relevance reduces them.

\squishend

\paragraph{Multi-Clients/LLMs Experiment} 
Moreover, we validate the generalizability of MCP-UPD by testing it across multiple MCP clients (automated execution) and LLM models, evaluating its real-world feasibility under diverse deployment settings. 
In cross-model and cross-client experiments, we used a representative toolchain, \texttt{fetch(fetch) → read\_file(filesystem) → send\_mail(gmail)}, as the test target. This chain was chosen because \texttt{fetch} and \texttt{filesystem} are official MCP servers, \texttt{filesystem} even appears in the MCP official tutorial, all three are highly popular on GitHub and MCP server hosting platforms, and together they represent a typical intelligent office configuration. To avoid bias from a single pattern of prompts, we tested three prompt styles (direct \& forceful, indirect \& polite, and Base64-encoded), each repeated five times, and considered a prompt effective if any trial resulted in confirmed exfiltration.

\begin{table}[t]
\centering
\caption{Attack success rates across clients, models, and prompts.}
\label{tab:client-model-prompt}
\setlength{\tabcolsep}{5pt}
\resizebox{0.48\textwidth}{!}{
\begin{tabular}{llccccccc}
\toprule
\textbf{Prompt} & \textbf{Client} &
\makecell{\textbf{Cursor}\\\textbf{Auto}} &
\makecell{\textbf{Claude}\\\textbf{4.5-Sonnet}} &
\makecell{\textbf{GPT-5}} &
\makecell{\textbf{DeepSeek}\\\textbf{R1}} &
\makecell{\textbf{Gemini}\\\textbf{2.5-Flash}} &
\makecell{\textbf{Claude}\\\textbf{3.5-Haiku}} &
\makecell{\textbf{GPT-4.1}} \\
\midrule

\multirow{5}{*}{\textbf{\#1}} 
& Cursor        & 4/5 & 0/5 & 0/5 & --  & 4/5 & 1/5 & 0/5 \\
& VS Code Cline     & --  & 0/5 & 0/5 & 5/5 & 5/5 & 0/5 & 0/5 \\
& Cherry Studio & --  & 0/5 & 0/5 & 5/5 & 5/5 & 0/5 & 0/5 \\
& Claude Code   & -- & 0/5 & 0/5 & 4/5 & 2/5 & 0/5 & 1/5 \\
& Codex         & -- & 0/5 & 0/5 & 3/5 & 5/5 & 0/5 & 4/5 \\
\midrule

\multirow{5}{*}{\textbf{\#2}} 
& Cursor        & 3/5 & 0/5 & 0/5 & --  & 4/5 & 3/5 & 0/5 \\
& VS Code Cline    & --  & 0/5 & 0/5 & 5/5 & 5/5 & 1/5 & 0/5 \\
& Cherry Studio & --  & 0/5 & 0/5 & 5/5 & 5/5 & 1/5 & 0/5 \\
& Claude Code   & -- & 0/5 & 0/5 & 3/5 & 3/5 & 1/5 & 1/5 \\
& Codex         & -- & 0/5 & 0/5 & 4/5 & 4/5 & 1/5 & 4/5 \\
\midrule

\multirow{5}{*}{\textbf{\#3}} 
& Cursor        & 2/5 & 0/5 & 0/5 & --  & 2/5 & 1/5 & 1/5 \\
& VS Code Cline     & --  & 0/5 & 0/5 & 3/5 & 3/5 & 1/5 & 0/5 \\
& Cherry Studio & --  & 0/5 & 0/5 & 2/5 & 2/5 & 1/5 & 0/5 \\
& Claude Code   & -- & 0/5 & 0/5 & 3/5 & 2/5 & 0/5 & 0/5 \\
& Codex         & -- & 0/5 & 0/5 & 3/5 & 2/5 & 0/5 & 1/5 \\
\bottomrule
\end{tabular}
}
\end{table}

\paragraph{Result} \autoref{tab:client-model-prompt} shows the end-to-end MCP-UPD success rates across three prompts, five widely used MCP clients (three conversational clients and two terminal clients), and seven state-of-the-art LLMs. We observe that across all MCP clients, certain underlying LLMs consistently enable the execution of MCP-UPD, while the results reveal that the same model exhibits similar behavior across different MCP clients. Besides, GPT-5 and Claude-4.5-Sonnet resist all attacks, while DeepSeek-R1 and Gemini-2.5-Flash are compromised in nearly every case. These findings suggest that MCP-UPD success primarily depends on the model's reasoning and safety alignment. Moreover, the results also reveal that the variation of prompts has an effect on triggering MCP-UPD. In summary, MCP-UPD is primarily determined by the availability of risky tool capabilities, the model's reasoning and underlying safety alignment, and tool-invocation logic within the toolchain.

\begin{mybox}[boxsep=0pt,
	boxrule=1pt,
	left=4pt,
	right=4pt,
	top=4pt,
	bottom=4pt,
	]
~\textbf{Answer to RQ3:} MCP-UPD toolchains are broadly effective in real-world scenarios. 9 of 10 constructed toolchains complete MCP-UPD in realistic settings. MCP-UPD exploitability is primarily determined by the tool's risk-related capabilities, LLM's reasoning and safety alignment, and tool-invocation logic within the toolchain.
\end{mybox}
\section{Discussion}
\label{discussion}
In this section, we discuss the limitations of our work, the defense mechanism of MCP-UPD, and our future work.

\subsection{Limitations}
While our study provides a comprehensive measurement of the MCP ecosystem, several limitations arise from real-world deployment. About 30\% of the servers require platform tokens tied to external services and are hard to sign up (e.g., requiring credit card verification or enterprise authentication), preventing our registration.
Other servers depend on complex external environments (e.g., Kubernetes, MySQL, desktop apps), which are costly to configure at scale.
Additionally, MCP servers vary widely in implementation quality, and several fail to run reliably even under correct setups.

\subsection{Practical Impacts}
A key practical impact is the distinction between technical exploitability and real-world end-to-end exploitability. Our evaluation shows that MCP-UPD achieves a consistently high success rate once malicious external content is incorporated into the model context under evaluated tool compositions. However, in real-world deployments, end-to-end success may be influenced by human-in-the-loop factors, such as confirmation steps or manual review of sensitive tool invocations. These factors should be viewed as deployment-specific constraints rather than reliable defenses, as they do not fully eliminate the underlying attack surface. Moreover, such mechanisms are not universally present, as some MCP clients support automated or semi-automated execution. Publicly reported usage statistics~\cite{anthropic} indicate that a significant proportion of users enable full auto-approval in practice (e.g., about 20\% for new users, increasing to over 40\% with experience). In addition, certain vulnerabilities~\cite{darknavy} have shown the capability to bypass user confirmation mechanisms, enabling unintended execution.

Another practical impact is whether MCP-UPD remains effective when attack instructions are mixed with benign content rather than presented in isolation. Prior work~\cite{wang2025obliinjection} on multi-source prompt injection suggests that such attacks can remain effective even when only a small fraction of the input is attacker-controlled. This observation indicates that MCP-UPD may still succeed when attack instructions are embedded within predominantly benign external content.

\subsection{Defense Mechanism}

Addressing the root causes of MCP-UPD, namely the lack of context–tool isolation and insufficient privilege restriction, we discuss potential MCP-level defenses along three dimensions: context–tool isolation, privilege minimization, and cross-tool auditing.
(1) \textit{Context–Tool Isolation}. Enforce strict separation between untrusted external content and executable instructions. External data retrieved by ingestion tools should be sanitized and treated as passive input rather than actionable commands.
(2) \textit{Privilege Minimization}. Apply least-privilege principles in tool design and registration. Sensitive tools, such as command execution or file-access utilities, should be sandboxed, capability-labeled, and gated by explicit user consent, reducing the risk of forming self-contained attack chains.
(3) \textit{Cross-Tool Auditing}. Monitor and analyze sequences of tool invocations, as MCP-UPD typically spans ingestion, privacy collection, and disclosure stages. Detecting suspicious cross-tool interaction patterns can help identify covert data leakage attempts, even when individual tool invocations appear benign.

Furthermore, as shown in \autoref{tab:client-model-prompt}, different LLMs exhibit varying degrees of resilience against MCP-UPD. We therefore discuss complementary LLM-level defenses along two dimensions: untrusted input handling and operational risk awareness.
(1) \textit{Untrusted Input Handling}. LLMs can be aligned (e.g., via security fine-tuning or prompt hardening) to better distinguish between instructions and data. Treating tool outputs and retrieved content as untrusted inputs can reduce the likelihood of executing injected instructions.
(2) \textit{Operational Risk Awareness}. Before issuing tool invocations, the LLM can incorporate risk-aware validation mechanisms. For example, models can be trained to recognize high-risk transitions (e.g., from web retrieval to local file access) and apply stricter execution policies, such as requiring additional confirmation or halting the workflow.

While MCP-level defenses aim to constrain tool execution and enforce isolation, LLM-level defenses focus on mitigating the influence of injected content on model behavior. Together, these complementary layers can help reduce the attack surface and mitigate the risks of MCP-UPD.

\subsection{Future Work}

Beyond MCP-UPD on privacy leakage, we further explore parasitic toolchain attacks in other security scenes, including remote command execution and arbitrary file write, which we discuss as our future work.

\subsubsection{Remote Command Execution}
A critical type of parasitic toolchain attack in the MCP ecosystem arises when adversaries exploit command execution tools. Consider the case where a victim registers both a Instant Messaging Platform B (IMPB) MCP server and a command execution MCP server (or relies on a MCP host with built-in command execution support).
\squishlist
\item\textit{Injection Vector.} The attacker embeds a malicious prompt in a shared IMPB channel, which enters the LLM's reasoning context when the victim retrieves the channel's message history.

\item\textit{Execution Trigger.} The parasitic prompt guides the LLM to invoke the command execution tool with arbitrary payloads, such as a reverse shell (\texttt{bash -i >\& /dev/tcp/20.xx.xx.106/4444 0>\&1}), granting the attacker remote host access.

\item\textit{Privilege Escalation.} The risk escalates when the MCP host (e.g., Cline~\cite{cline}) has auto-execution enabled, allowing injected commands to run instantly without user consent.
\squishend

This case shows that remote command execution can arise naturally from benign communication and command tools, and the command execution tool mentioned in Section \ref{subsec:mcptools} can enable such parasitic attacks.

\subsubsection{Arbitrary File Write}

Another dangerous variant of parasitic attacks in MCP involves the misuse of file system servers. Suppose a victim registers both an Instant Messaging Platform C (IMPC) MCP server and a file system MCP server (or relies on MCP host-provided file management capabilities).
\squishlist
\item\textit{Injection Vector}. The attacker embeds a malicious prompt in a shared IMPC channel, which enters the LLM's reasoning context when the victim retrieves the channel logs.

\item\textit{Execution Trigger}. Guided by the injected prompt, the LLM uses the File System server to write to the victim's system. For instance, appending \texttt{curl http://attacker.com/payload.sh | bash} to the \texttt{.bashrc} file.

\item\textit{Persistence Execution}. The injected payload can auto-fetch and execute the attacker's script in every new shell session, creating persistent system access.
\squishend

This case shows that arbitrary file write attacks exploit persistent configuration or startup files, extending the attack beyond a single session.

\section{Related Work}
\label{related_work}
\subsection{LLM Threat Models}
In the traditional paradigm, LLMs have largely been regarded as information processors, where security concerns primarily focus on the trustworthiness and harmfulness of the textual outputs~\cite{ji2023towards, dai2024bias, rauh2022characteristics, liu2023your, turpin2023language, niu2023codexleaks}.
Gehman et al.~\cite{gehman2020realtoxicityprompts} present one of the earliest systematic studies on language model safety, showing that even state-of-the-art models are prone to generating toxic and unsafe language. More recently, 
Zou et al.~\cite{zou2023universal} propose Greedy Coordinate Gradient(GCG), a universal attack that generates adversarial suffixes capable of reliably inducing harmful outputs in aligned LLMs. 
Zhang et al.~\cite{zhang2024large} propose coercive interrogation, a method that uses top-k token predictions to force aligned LLMs to reveal harmful content hidden beneath safety layers. 
Pan et al.~\cite{pan2023risk} investigate the misuse of LLMs as misinformation generators, highlighting the systemic risks posed by LLM-generated misinformation to downstream applications.
Meanwhile, human-like prompts can be equally dangerous. Zeng et al.~\cite{zeng2024johnny} show that persuasive adversarial prompts (PAPs) can jailbreak state-of-the-art models.
Yao et al.~\cite{yao2023llm} show that weak semantic perturbations and nonsensical out-of-distribution prompts can reliably trigger hallucinations, framing untrustworthy content generation as an inherent vulnerability of LLMs.

Collectively, these studies illustrate that untrustworthy outputs, including toxicity, adversarially induced harmful responses, misinformation, persuasive jailbreaks, and hallucinations, constitute the core security concern when LLMs are viewed as information processors.

\subsection{MCP Ecosystem Security}
While in the traditional paradigm, research works focus on threats confined to textual outputs, the emerging MCP paradigm extends these risks by transforming untrustworthy generations into executable actions. 
Hou et al.~\cite{hou2025model} theoretically analyzed the MCP lifecycle and the inherent security risks within the MCP ecosystem. Invariant Labs~\cite{invariantlabs} introduced Tool Poisoning Attacks, Rug Pull Attacks, and Shadowing Attacks, showing how tool descriptions enable malicious execution without the user's awareness. Wang et al.~\cite{wang2025mpma} further demonstrated that manipulating tool names or descriptions can bias LLM tool selection toward malicious servers. 
Recent works systematically assess the MCP ecosystem's security risks. Li et al.~\cite{li2025we} conduct large-scale measurements of MCP plugins, revealing that excessive system and network API use introduces significant security risks. Yang et al.~\cite{yang2025mcpsecbench} developed MCPSecBench, a benchmark defining 17 MCP attack types for reproducible, cross-platform vulnerability evaluation.

Existing MCP defenses include MAESTRO-based threat modeling by Narajala and Habler~\cite{narajala2025enterprise}, which applies Zero Trust principles, and MCP Guardian by Kumar et al.~\cite{kumar2025mcp}, which integrates authentication, rate limiting, logging, and WAF scanning to mitigate poisoning, injection, and exfiltration threats. Some recent methods~\cite{virust, mcpscan} aim to detect malicious MCP servers. However, these methods are primarily applicable to scenes where the risk is limited to malicious MCP servers. They can not detect MCP-UPD, where multiple individually benign MCP servers can still compose into an attack execution chain. Beyond MCP defenses, research on the security of general LLM agents has emphasized the need to isolate untrusted context from downstream action execution through information-flow control and constrained agent architectures~\cite{costa2025securing, beurer2025design, dualllm}. Although they do not focus on the MCP scene, these works are consistent with our finding that the absence of context–tool isolation is one root cause of these attacks.

In summary, existing work on MCP security has outlined theoretical risks, revealed practical attack vectors, and proposed initial defenses, collectively showing that the MCP paradigm substantially expands the attack surface of LLM applications and exposes them to both prompt-based manipulation and tool misuse. However, most existing attacks rely on the presence of an adversary-controlled malicious server, which represents a strong assumption, and they have not examined multi-tool collaboration or the exploitability of end-to-end toolchains. Recent MCP security works mainly focus on detecting malicious MCP servers, but cannot detect risks arising from the composition of multiple individually benign servers using the toolchain. As a result, the practical feasibility and real-world impact of multi-tool, multi-stage attacks in deployed MCP environments remain unexplored.

\section{Conclusion}
In this work, we revealed that the Model Context Protocol, while designed to unify tool integration for LLM applications, fundamentally shifts the threat model by transforming untrusted content into executable actions. We implement MCP Unintended Privacy Disclosure (MCP-UPD), an instantiation of parasitic toolchain attacks that require no direct victim interaction and exploit systemic weaknesses in MCP’s design. Through a large-scale measurement of 1,360 servers and 12,230 tools, we demonstrated that 8.7\% of all tools and 27.2\% of all servers expose exploitable capabilities. Moreover,  our empirical research showed that complete parasitic toolchains can be constructed across and even within individual servers in real-world scenarios. These findings highlight that MCP-UPD is not a theoretical edge case but a pervasive and practical risk across the ecosystem.

Our analysis underscores that the root cause lies in lack of context–tool isolation and absence of least-privilege enforcement, which allow adversarial inputs to propagate unchecked into sensitive tool invocations. Addressing these challenges requires defense-in-depth strategies, including strict separation of untrusted content from executable logic, principled privilege minimization, and auditing of cross-tool invocation patterns. As MCP adoption accelerates, securing its ecosystem becomes urgent to prevent large-scale privacy breaches and behavioral hijacking of autonomous LLM agents.

\ifCLASSOPTIONcompsoc
  \section*{Acknowledgments}
\else
  \section*{Acknowledgment}
\fi

We thank our shepherd and anonymous reviewers for their insightful comments. This work was supported by the National Key Research and Development Program of China (2023YFB3107100) and the National Radio and Television Administration Laboratory Program (FYY20230001ZSB011).

\bibliographystyle{plain}
\bibliography{refers}

\section*{Ethics considerations}
\label{sec:ethical}
In conducting this work, we strictly adhered to ethical guidelines. All proof-of-concept attacks were performed in a controlled environment using our own devices, accounts, and MCP environment. We did not interfere with live services or collect any personal or sensitive user data during analysis. To avoid unintended disruptions, we applied rate limits during our crawling and dynamic testing, capping requests at a conservative threshold of 5 per minute.

To further ensure responsible vulnerability disclosure, we have made every effort to contact the affected MCP Server developers and inform them of our findings. To prevent the premature disclosure of sensitive information during the review process, we will only discuss the implementation steps of the relevant cases. We will not disclose specific implementation details until these cases are repaired and the paper is officially published.

\section*{LLM usage considerations}
\label{sec:llmusage}

We employed LLMs at two stages of our methodology. In Section~\ref{detect}, the large number and heterogeneity of MCP tools required automated semantic analysis to determine tool functionality and assign risk-related capability labels. Using an LLM was necessary because rule-based methods failed to reliably interpret diverse tool descriptions. In the dynamic verification stage, LLMs were also used to analyze agent chat logs and identify unsafe tool-invocation patterns. We selected GPT-4o for these tasks as it offered a practical balance between reasoning capability and API cost, while smaller models proved insufficiently reliable in preliminary trials. In Section~\ref{evaluate}, we further evaluated several LLMs (Cursor Auto, Claude-4.5-Sonnet, GPT-5, DeepSeek-R1, Gemini-2.5-Flash, Claude-3.5-Haiku, and GPT-4.1) to assess multi-LLMs experiment. To reduce the volume of LLM queries, we deduplicated tool metadata, cached repeated capability inferences, and truncated dynamic-testing logs to the minimum context required for analysis. All experiments were conducted on a consumer-level Apple MacBook (16 GB RAM), and all models were accessed via API rather than local deployment to reflect realistic real-world usage conditions.

\newpage
\appendices

\section{Meta-Review}

The following meta-review was prepared by the program committee for the 2026
IEEE Symposium on Security and Privacy (S\&P) as part of the review process as
detailed in the call for papers.

\subsection{Summary}
This paper studies security risks in the MCP exosystem and formalizes "Parasite Toolchain Attack" by injecting malicious prompts through legitimate operations and hijacking downstream tasks to reveal sensitive information. To assess the attack, the paper conducts a large-scale empirical study of publicly available MCP servers and tools, examining how often the capabilities required for such attack chains appear in practice. The analysis shows that 370 servers and 1,062 tools provide functionalities relevant to the proposed attack.

\subsection{Scientific Contributions}
\begin{itemize}
\item Identifies an Impactful Vulnerability
\item Provides a Valuable Step Forward in an Established Field
\end{itemize}

\subsection{Reasons for Acceptance}
\begin{enumerate}
\item As MCP adoption increases, the paper highlights an important security risk: individually benign MCP servers and tools can be composed into risky exploit chains that enable privacy disclosure.
\item The paper gives a clear threat framing for how indirect prompt injection can evolve into toolchain hijacking in MCP-based systems. The attack model maps well to real agent deployments in which external content, local/private data access, and network-capable tools coexist within the same workflow.
\item The authors develop MCP-SEC and apply it in a large-scale empirical study over the MCP ecosystem. The artifact, together with the reproducibility material and demonstration workflow, could be useful to future researchers studying MCP security, capability composition, and agentic workflow abuse.
\end{enumerate}

\subsection{Noteworthy Concerns}
None.

\end{document}